\documentclass[5p,twocolumn,sort&compress,times,pdftex,preprint]{elsarticle}

\usepackage{amsmath, amsfonts, amsthm, amssymb}
\usepackage{color}
\usepackage{url}
\usepackage{ulem}
\usepackage{bm}
\usepackage{bbold}
\usepackage{amsmath}
\usepackage{algorithm}
\usepackage{algpseudocode}
\usepackage{braket}
\usepackage{multirow}

\definecolor{myorange}{rgb}{0.7,0.5,0.0}
\definecolor{mygreen}{rgb}{0.2,0.6,0.0}
\def\vec#1{\boldsymbol #1}

\newcommand{\tohgoe}[1]{\textcolor{black}{#1}}

\newcommand{\tido}[1]{\textcolor{black}{#1}}
\newcommand{\tmorita}[1]{\textcolor{black}{#1}}
\newcommand{\kato}[1]{\textcolor{black}{#1}}
\newcommand{\yoshimi}[1]{\textcolor{black}{#1}}
\newcommand{\imada}[1]{\textcolor{black}{#1}}

\newcommand{\tg}[1]{\textcolor{black}{#1}}

\newcommand{\Ns}{N_{\rm s}}
\newcommand{\Ne}{N_{\rm e}}

\newcounter{bla}

\journal{Computer Physics Communications}

\begin{document}

\begin{frontmatter}



\title{mVMC -- Open-source software for many-variable variational Monte Carlo method}


\author[a]{Takahiro Misawa\corref{author}}
\author[a]{Satoshi Morita}
\author[a]{Kazuyoshi Yoshimi}
\author[a]{Mitsuaki Kawamura}
\author[a]{Yuichi Motoyama}
\author[b]{Kota Ido}
\author[b]{Takahiro Ohgoe}
\author[b]{Masatoshi Imada}
\author[a]{Takeo Kato}

\cortext[author] {Corresponding author.\\\textit{E-mail address:} tmisawa@issp.u-tokyo.ac.jp}
\address[a]{Institute for Solid State Physics, University of Tokyo, Kashiwa-shi, Chiba, 277-8581,Japan}
\address[b]{Department of Applied Physics, University of Tokyo, Bunkyo-ku, Tokyo, 113-8656, Japan}

\begin{abstract}
mVMC (many-variable Variational Monte Carlo) is an open-source software 
based on the variational Monte Carlo method applicable for 
a wide range of 
{Hamiltonians} for interacting fermion systems.
In mVMC, we introduce more than ten thousands variational parameters
and simultaneously optimize them by using the 
stochastic reconfiguration (SR) method.
In this paper, we explain basics and user interfaces of mVMC.
By using mVMC, users can perform the calculation by preparing only one input file of about ten lines
for widely studied quantum lattice models, and can also perform it for general
Hamiltonians by preparing several additional input files.
We show the benchmark results of mVMC for the Hubbard model, 
the Heisenberg model, and the Kondo-lattice model. 
These benchmark results demonstrate that mVMC 
{provides}
ground-state and low-energy-excited-state wave functions for interacting fermion systems
with high accuracy.

\end{abstract}

\begin{keyword}
02.60.Dc Numerical linear algebra\sep
71.10.Fd Lattice fermion models
\end{keyword}

\end{frontmatter}



{\bf PROGRAM SUMMARY}

\begin{small}
\noindent
{\em Manuscript Title:} mVMC -- Open-source software for many-variable variational Monte Carlo method \\
{\em Authors:}  
Takahiro Misawa, Satoshi Morita, Kazuyoshi Yoshimi, Mitsuaki Kawamura, Yuichi Motoyama, Kota Ido, Takahiro Ohgoe, Masatoshi Imada, Takeo Kato \\
{\em Program Title:} mVMC \\
{\em Journal Reference:}   \\
{\em Catalogue identifier:}                                   \\
{\em Program summary URL:} \\
http://ma.cms-initiative.jp/en/application-list/mvmc \\
{\em Licensing provisions:} GNU General Public License, version 3 or later\\
{\em Programming language:} C                                   \\
{\em Computer:} Any architecture with suitable compilers including PCs and clusters.\\
{\em Operating system:} Unix, Linux, OS X.  \\
{\em RAM:} Variable, depending on number of variational parameters. \\
{\em Number of processors used:} Arbitrary (Monte Carlo samplings are parallelized by using MPI). \\
{\em Keywords:}
02.70.Ss Quantum Monte Carlo methods,
71.10.Fd Lattice fermion models
\\
{\em Classification:}
4.12 Other Numerical Methods, 7.3 Electronic Structure
\\
{\em External routines/libraries:} {MPI, BLAS, LAPACK, ScaLAPACK (optional)}\\
{\em Nature of problem:}\\
Physical properties (such as the charge/spin structure factors)
of strongly correlated electrons at zero temperature.
\\
{\em Solution method:}\\
Application software based on the variational Monte Carlo method 
for quantum lattice model such as the Hubbard model, 
the Heisenberg model and the Kondo model.
\\
{\em Unusual features:}\\
It is possible to perform the highly-accurate calculations for ground states in a wide range of 
theoretical Hamiltonians in quantum many-body systems. In addition to the 
conventional orders such as magnetic and/or charge orders, 
user can treat the anisotropic superconductivities 
within the same framework. This flexibility is the main advantage of mVMC.
\\
\\

\end{small}


\section{Introduction}
\label{Intro}

High-accuracy analyses of effective Hamiltonians for interacting fermion systems have been an important issue
for a long time in {studies} of novel quantum phases 
in strongly correlated electron systems such 
as high-temperature {superconductors}~\cite{Imada_RMP1998} 
and quantum spin liquids~\cite{Diep,Balents_Nature2010}. 
Recent theoretical development in construction of low-energy effective Hamiltonians 
for real materials in a non-empirical way~\cite{Imada_JPSJ2010} 
{urges developing ways for}
high-accuracy 
analyses of 
{Hamiltonians} with complex hopping parameters and 
electron-electron interactions.
To promote materials design of correlated electron systems, 
it is highly 
{desirable} to develop
a numerical solver, which can analyze a wide range of 
{Hamiltonians} with high accuracy.

To solve 
{Hamiltonians} for interacting fermion systems,
various numerical algorithms have been developed so far~\cite{FehskeBook2008}. The exact diagonalization method is one
of {most} reliable theoretical methods applicable to general 
{Hamiltonians}~\cite{Dagotto_RMP1994,hphi}.
However, system sizes which can be treated in the exact diagonalization method
are limited, because the Hilbert-space dimensions 
{increase} exponentially as a function of systems sizes.
The (unbiased) quantum Monte 
{Carlo} method is another important method applicable to
a wide range of quantum many-body systems~\cite{KawashimaBook}.
For interacting fermion systems, however, the negative sign problem, i.e., the appearance of
the negative weights in the Monte Carlo samplings makes it {difficult} 
or 
{practically} impossible to obtain reliable results for 
a realistic numerical cost {except {a} few special cases}.
For one-dimensional quantum systems, the density matrix renormalization group (DMRG)
is an excellent method~\cite{White_PRL1992,White_PRB1993,Schollwoeck_RMP2005}, which can treat larger system sizes.
The tensor-network method, which has been developed keeping close relation with the DMRG method,
{has succeeded in treating two-dimensional systems 
without suffering {from} the negative sign problem~\cite{Cirac_JPhysA2009,Orus_AnnPhys2014,Ran_arXiv2017}.}
{
However, the computational time by the tensor network 
method increases very rapidly with the increasing 
tensor dimension and the convergence to
the exact estimate is not throughly examined 
so far in complex systems. Particularly in cases of 
itinerant fermion models we need special care 
about the entanglement entropy, which increases beyond 
the area law and the accuracy by the tensor network becomes worse.
}

The variational Monte Carlo (VMC) method is one of promising methods
{for} 
highly 
accurate calculations for 
{general systems}~\cite{Gros_AP1989}.
In the VMC \yoshimi{calculation}, we introduce a variational wave function with variational parameters,
and obtain approximate ground-state wave functions by optimizing these parameters
according to the variational principle. For 
{calculating} expectation values of 
physical quantities, we employ the Markov-chain Monte Carlo sampling.
In contrast to the ordinary (unbiased) Monte Carlo method,
the VMC \yoshimi{method} does not suffer {from} the negative sign problem since the weight of
Monte Carlo sampling is positive definite.
The VMC \yoshimi{method} was applied to various quantum many-body systems such
as liquid ${}^4$He~\cite{McMillan1965}, liquid ${}^3$He~\cite{Ceperley1977},
the Hubbard model~\cite{Gros1987,Yokoyama1987,Giamarchi1991,Eichenberger2007,Tocchio2008,Yokoyama2013,Tocchio2014}, 
the Kondo-lattice model~\cite{Watanabe2007,Asadzadeh2013}, and the Heisenberg model~\cite{Liang1990,Liang1990a,Franjic1997}.
In these applications, trial wave functions were assumed to be a simple mean-field form with only a few variational parameters,
and were optimized to reproduce ground-state properties qualitatively.
\kato{
The accuracy of the VMC calculation using such simple trial wave functions is, however, not satisfactory, 
if one hope to identify a novel quantum phase competing with other possible phases, 
or to treat complicated low-energy Hamiltonians for real materials.} 

To improve the accuracy of the VMC {method} substantially, 
more than ten thousand variational parameters 
{are introduced} 
in trial wave functions 
{where they are}
simultaneously {optimized} 
by using the 
stochastic reconfiguration (SR) method~\cite{Sorella_PRB2001,Sorella_JCS2007}.  
In addition, we implement quantum number projections to restore the symmetries of the wave functions 
to improve 
{their} accuracy. 
{We have developed a program package named
mVMC (many-variable Variational Monte Carlo), which
can perform highly accurate VMC calculations
for a wide range of the quantum lattice models.}
mVMC 
has been applied to the Hubbard models~\cite{Tahara_JPSJ2008,Tahara_JPSJLett2008,Misawa_PRB2014}, 
the Heisenberg models~\cite{Kaneko_JPSJ2014,Morita_JPSJ2015}, and 
the Kondo-lattice models~\cite{Motome_PRL2010,Misawa_PRL2013}, and has succeeded in
evaluating electronic states with high accuracy.
mVMC 
has also applied to more realistic models 
such as the theoretical models for the 
interfaces of the cuprates (stacked Hubbard models)~\cite{Misawa_SA2016},
{\it ab initio}
low-energy Hamiltonians for the 
iron-based superconductors~\cite{Misawa_JPSJ2011,Misawa_PRL2012,Misawa_Ncom2014,Hirayama_JPSJ2015}, 
and 
{\it ab initio}
low-energy Hamiltonians for the organic conductor~\cite{Shinaoka_JPSJ2012}.
It has been shown that the mVMC 
can be applied to 
systems with spin-orbit couplings~\cite{Yamaji_PRB2011,Kurita_PRB2015,Kurita_PRB2016} 
and 
systems with 
electron-phonon couplings~\cite{Ohgoe_PRB2014,Ohgoe_PRL2017}.
Recently, it has been shown that 
mVMC 
can
{treat} the real-time evolutions~\cite{Ido_PRB2015,Ido_2017} 
and {perform} 
finite-temperature calculations~\cite{Takai_JPSJ2016} 
based on the imaginary-time evolutions.

Recently, the authors have 
{released} 
mVMC as open-source software 
with simple and flexible user interfaces~\cite{ma}.
\kato{By using this software, users can perform many-variable VMC calculations
for widely studied quantum lattice models by preparing only one input file with less than ten lines.}
{Although there are several open-source software of VMC method for continuous
space such as CASINO~\cite{CASINO}, QWalk~\cite{QWalk}, and turbo-RVB~\cite{TRVB}, there is no open-source 
software of VMC method for the quantum lattice model such as the 
Hubbard or the Heisenberg model to the best of our knowledge.}
By preparing several additional input files, users can also define general Hamiltonians 
with \imada{any lattice structure, any spatial dimensions and any one and two-body (interaction) terms.}
The user interface of mVMC is designed for seamless connection to open-source
software $\mathcal{H}\Phi$~\cite{hphi,hphi_ma}, which is developed by some of the authors
for 
exact diagonalization calculations.
By small changes of description in input files, it is easy to check the accuracy of the 
variational wave functions by comparing the results 
{to} the exact diagonalization calculations for small-size systems.
mVMC also supports 
large-scale parallelization. 
\tg{In mVMC,} the power-Lanczos 
\tg{method}\tg{\cite{Heeb_ZPhys1993}} \tg{is implemented.} 
\tg{This method} systematically improve\tg{s} the accuracy 
of the wave functions\tg{.} 

In this paper, we describe basic usage of mVMC
and fundamental properties of trial wave functions implemented in mVMC. 
We also 
{exposit} the key algorithms implemented 
in mVMC such as the quantum number projections and the SR method.
We show some examples of mVMC calculations and
{demonstrate that the relative {systematic} errors 
{on} the energy 
become typically less than $10^{-2}\%$
compared with the results of the exact diagonalization method.
}

\begin{figure}[tb!]
  \begin{center}
    \includegraphics[width=9.5cm]{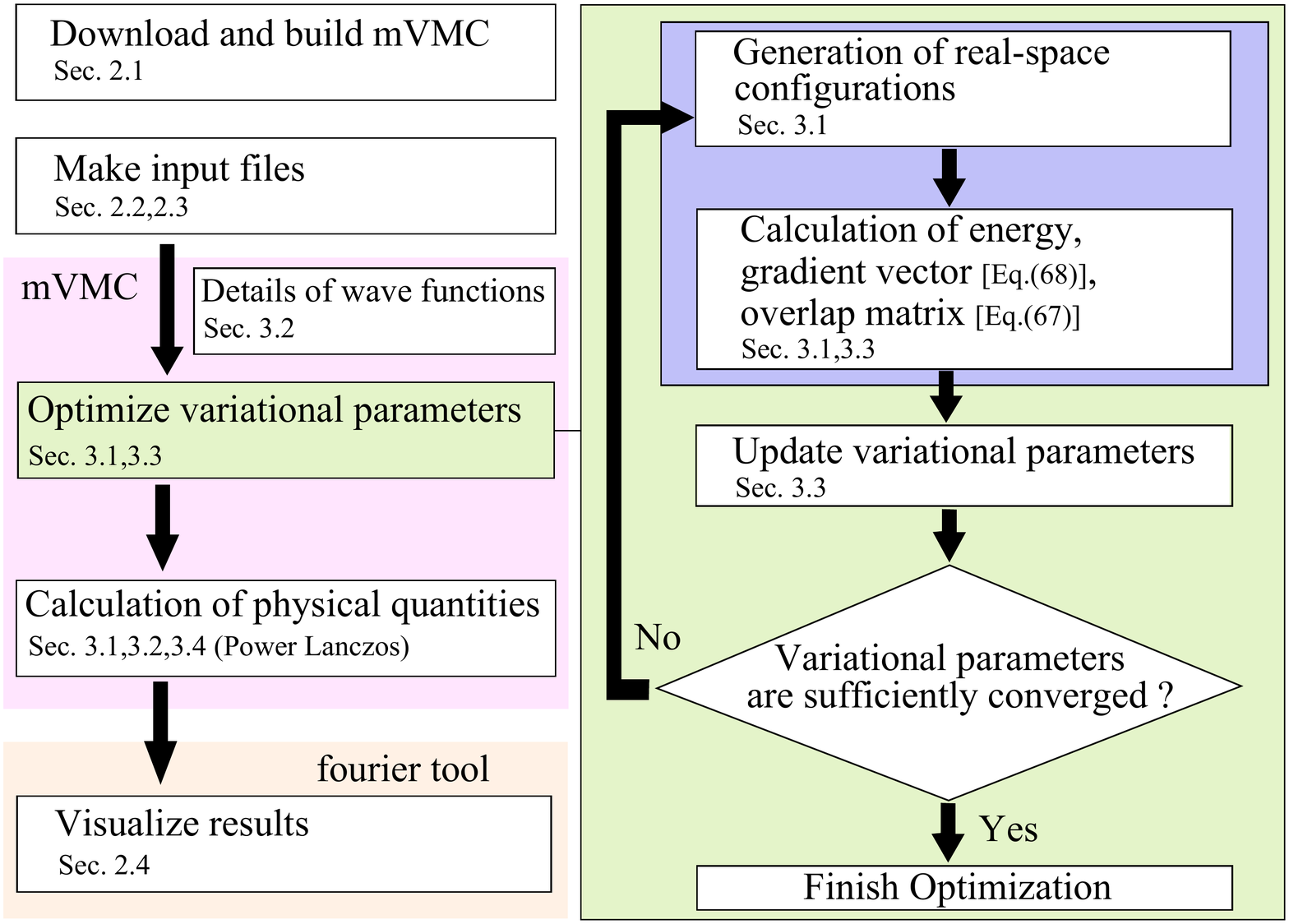}
    \caption{
	\yoshimi{Overview of this paper. 
        For readers who want to use mVMC quickly,
        please read Section 2 first, where we describe basic usage of mVMC. 
        Section 3 \imada{orange} be useful for readers who want to
        learn the key algorithms implemented in mVMC. }}
    \label{fig:overview}
  \end{center}
\end{figure}

This paper is organized as follows \yoshimi{(overview is shown in Fig. \ref{fig:overview})}:
In Section 2, we 
{describe} how to download and build mVMC (Section 2.1), 
how to use mVMC (Section 2.2), how to define models and lattices (Section 2.3),
and how to visualize results of mVMC calculation (Section 2.4).
In Section 3, we explain the algorithms implemented in mVMC.
We describe the {Monte Carlo} sampling method (Section 3.1), 
details of trial wave functions (Section 3.2),
the optimization method (Section 3.3), the power-Lanczos method (Section 3.4), and
the parallelizations (Section 3.5).
In Section 4, we show benchmark results of mVMC for the Hubbard model, the Heisenberg model,
and the Kondo-lattice model.
By these benchmarks, we demonstrate excellent performance of mVMC 
as a numerical solver for obtaining ground states and low-lying excited states 
of {the standard} Hamiltonians for interacting fermion systems.
Finally, Section 5 is devoted to the summary. 

\section{Basic usage of mVMC}
\subsection{How to download and build mVMC}
One can download the gzipped tar file of source codes~\cite{ma},
samples, and manuals from the mVMC download site.
It is also possible to download
the repository of mVMC through GitHub~\cite{git}.
For building mVMC,
a C 
{compiler}, the BLAS/LAPACK library~\cite{lapack}, 
and the Message Passing Interface (MPI)~\cite{MPI} are prerequisite.
The Scalapack~\cite{scalapack} library is optionally required.
By using the CMake utility~\cite{cmake}, one can build mVMC as follow:
\begin{verbatim}
$ tar xzvf mVMC-1.0.2.tar.gz
$ cmake mVMC-1.0.2/
$ make
\end{verbatim}
One can also select the compiler 
as follows:
\begin{verbatim}
$ cmake -DCONFIG=$Config $PathTomvmc
$ make
\end{verbatim}
where \verb|$Config| is chosen from the following configurations:
\begin{itemize}
\item \verb|gcc|     :  \yoshimi{GNU C Compiler}
\item \verb|intel|   : \yoshimi{Intel\textregistered~C Compiler} + MKL library
\item \verb|sekirei| : \yoshimi{Intel\textregistered~C Compiler} + MKL library on ISSP system-B (Sekirei)
\item \verb|fujitsu| : \yoshimi{Fujitsu C compiler} + SSL2 library on the K computer
\end{itemize}

We recommend users to use the CMake utility,
because the CMake utility automatically finds the required libraries.
However, installing CMake utility is sometimes difficult, for example,
in the systems where one does not have the administrative permission. 
Thus, for a system 
without the CMake utility, 
we provide a script 
for making the {Makefiles}. 
By using the  \verb|mVMCconfig.sh|,
one can generate the {Makefiles} as follows:
\begin{verbatim}
$ bash mVMCconfig.sh gcc-openmpi
$ make mVMC
\end{verbatim}
{This \tmorita{is} an example for gcc + openmpi configurations and
we provide several options of the combinations {between}
{compilers} and implementations of the MPI libraries
such as the intel compiler and mpich. For details,
please refer \tido{to} the manuals~\cite{ma}.}
Once the compilation finishes successfully, one can find the executable file,
\verb|vmc.out|, in \verb|src/mVMC/| subdirectory.

{
Another way to use mVMC is using MateriApps LIVE!~\cite{MALive}, which offers
an environment based on Debian GNU/Linux OS. 
MateriApps LIVE!
includes the standard libraries used in
computational materials science applications.
In MateriApps LIVE!, mVMC is pre-installed and 
user can use mVMC without compiling. 
We note that
MateriApps Installer~\cite{MAInstall} offers
scripts for installing mVMC in several different environments 
such as the supercomputers in Japan.}

\subsection{How to use mVMC}
\subsubsection{Expert mode}
To use mVMC, it is necessary to prepare several input files that 
{specify}
parameters in models, 
{forms of} the variational wave functions, 
and basic parameters for variational Monte Carlo calculations
such as the number of the Monte Carlo samplings.
In the list 
\yoshimi{file} (\verb|namelist.def| is a typical name of the list file), 
by using the keywords,
one can specify the types of 
input files.
Here, we show an example of the list of 
input files
below:
\begin{verbatim}
ModPara      modpara.def
LocSpin      locspn.def
Trans        trans.def
CoulombIntra coulombintra.def
OneBodyG     greenone.def
TwoBodyG     greentwo.def
Gutzwiller   gutzwilleridx.def
Jastrow      jastrowidx.def
Orbital      orbitalidx.def
TransSym     qptransidx.def
\end{verbatim}
For example, 
\verb|modpara.def| associated with the
keyword \verb|ModPara| is the input file for specifying the basic parameters for calculations,
and 
\verb|trans.def| associated with the keyword \verb|Trans| is the input file for specifying the 
transfer integrals in the model Hamiltonian.
In 
\tido{T}able \ref{table:Keywords}, we list keywords used in mVMC ver. 1.0.
\begin{table}[tb!]
\caption{Keywords used in mVMC. 
\yoshimi{By adding IN in front of the
keyword for variational parameters (Orbital, APOrbital, POrbital, GeneralOrbital,
Gutzwiller , Jastrow , DH2, and DH4) such as InOrbital}, 
one can specify the initial variational parameters.}
\begin{tabular}{ll}
\hline
keyword & explanation \\
\hline \hline
ModPara           & basic parameters for calculations\\ \hline \hline
LocSpn            & locations of local spins \\
Trans             & transfer integrals~(${\cal H}_T$) \\ 
CoulombIntra      & on-site Coulomb interactions~(${\cal H}_U$) \\  
CoulombInter      & off-site Coulomb interactions~(${\cal H}_V$) \\ 
Hund              & Ising-type Hund's rule couplings~(${\cal H}_H$) \\ 
Exchange          & exchange interactions~(${\cal H}_E$) \\ 
PairHop           & pair hopping terms~(${\cal H}_P$) \\ 
InterAll          & general two-body interactions~({${\cal H}_{\cal I}$}) \\   \hline \hline
Orbital/APOrbital & anti-parallel Pfaffian wave functions \\ 
POrbital          & parallel Pfaffian wave functions \\ 
GeneralOrbital    & general Pfaffian wave functions \\ 
Gutzwiller        & Gutzwiller correlations factors \\ 
Jastrow           & Jastrow    correlations factors \\   
DH2               & two-site doublon-holon correlations factors \\   
DH4               & four-site doublon-holon  correlations factors \\   
TransSym          & momentum and point-group projections \\   \hline \hline
OneBodyG          & one-body Green functions \\  
TwoBodyG          & two-body Green functions \\  
\hline
\end{tabular}
\label{table:Keywords}
\end{table}
After preparing all the necessary files,
one can start mVMC calculations by \yoshimi{executing} the following command:
\begin{verbatim}
$ ./vmc.out -e namelist.def
\end{verbatim}
In this procedure, one should prepare all the necessary files
correctly and it is sometimes 
{time consuming}.
To reduce efforts in preparing the input files,
we provide {a} simple mode called Standard mode 
for the standard models in the condensed matter physics
as explained {in} \tido{the} next subsection.

\subsubsection{Standard mode}
In Standard mode,
from the one input file \verb|StdFace.def|, 
all the necessary files are automatically generated.
By using Standard mode, one can treat 
the standard models in the condensed-matter physics
such as the Hubbard model, the Heisenberg model, 
the Kondo-lattice model, and their extensions.
In the following, we show an example  of the input file 
in Standard mode for
the Hubbard model on the square lattice:
\begin{verbatim}
model       = "FermionHubbard"
lattice     = "square"
W           = 4
L           = 4
Wsub        = 2
Lsub        = 2
t           = 1.0
U           = 4.0
nelec       = 16
2Sz         = 0
\end{verbatim}
Here, \verb|W| (\verb|L|) represents the
length for $x$ ($y$) direction
on the square lattice. 
\verb|Wsub| (\verb|Lsub|) is the length of
the sublattice structure {in the real space} for variational parameters.
\verb|model|
~\tido{and}  \verb|lattice| specify the types of the standard models
and lattice structures, respectively.
{The } hopping transfer and on-site Coulomb 
{interaction}
are represented by {\verb|t| and \verb|U|}, respectively.

By using this input file for the Hubbard model,
one can perform mVMC calculations by executing the following command.
\begin{verbatim}
$ ./vmc.out -s Stdface.def
\end{verbatim}
If one wants to check the input files without the calculation,
it is possible to generate only the input files without calculations by using the following command.
\begin{verbatim}
$ ./vmcdry.out Stdface.def
\end{verbatim}
Here, \verb|vmcdry.out| is the 
{executable} file
only for generating the input files. 

\subsubsection{Flow of calculation}
\begin{figure}[tb!]
  \begin{center}
    \includegraphics[width=9cm]{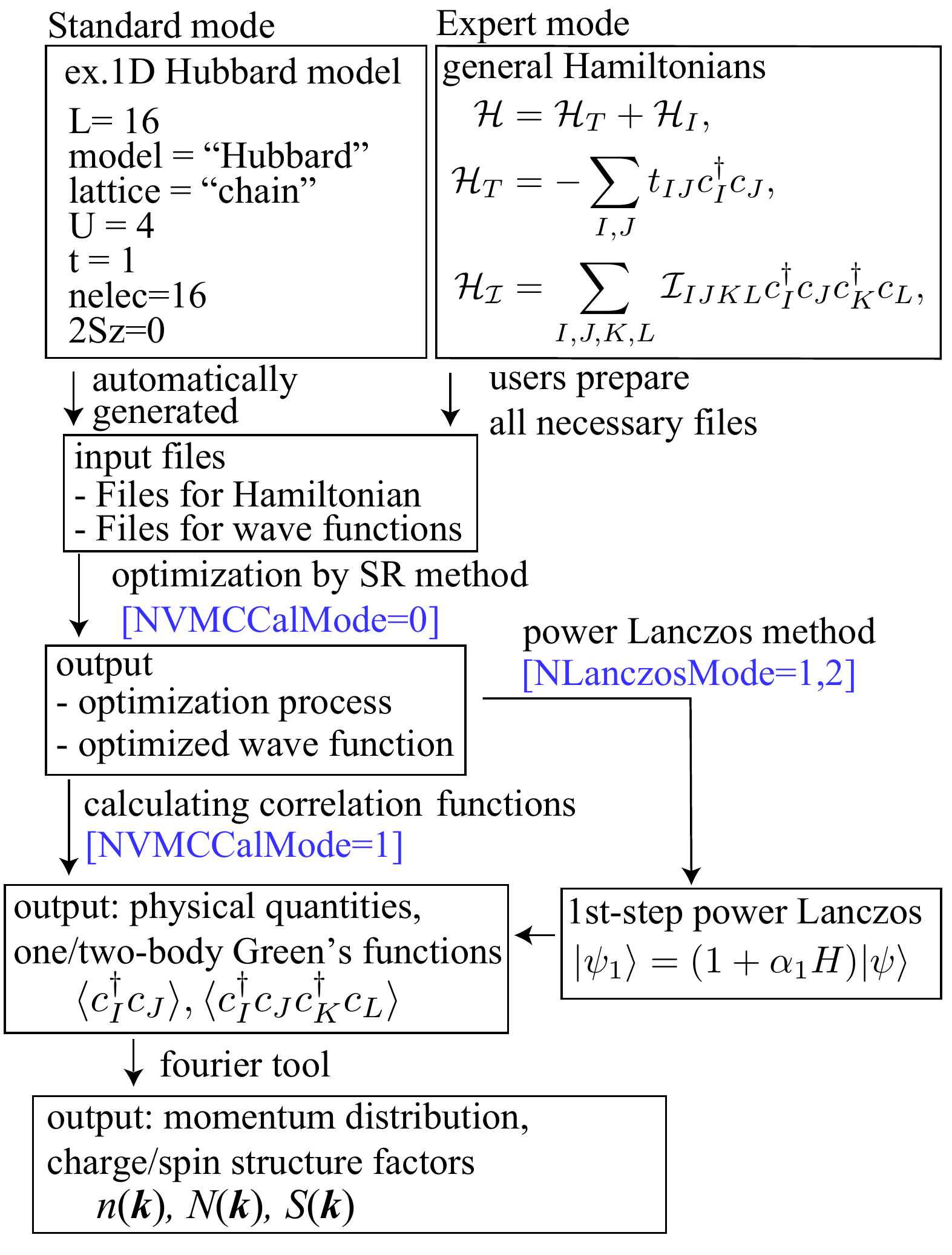}
    \caption{Flow of mVMC calculations. 
    {In the definitions of the Hamiltonians and the Green's functions,
    capital characters $I$,$J$,$K$,$L$ denote the site indices including the spin degreed of freedom, i.e.,
    $I=(i,\sigma_{i})$.}} 
    \label{fig:flow}
  \end{center}
\end{figure}

Here, we summarize the flow of mVMC calculations.
First, 
{one prepares}
\verb|StdFace.def| in Standard mode or 
all necessary input files in Expert mode.
Then, by executing \verb|vmc.out|,
{one performs}
the optimization of the variational parameters 
to lower the energy
according to the variational principle.
In the actual calculations, 
{one uses} 
the SR method~{\cite{Sorella_PRB2001,Sorella_JCS2007}}.
This optimization is the main part and the most time-consuming part of mVMC.
After the optimization,
by using the optimized variational wave functions,
{one calculates}
the specified 
correlation functions. 
This flow is summarized in Fig.~\ref{fig:flow}.
Optimization is performed for \verb|NVMCCalmode=0|
and the calculating physical 
\tido{quantities} 
for \verb|NVMCCalmode=1|. 
By choosing \verb|NLanczosMode=1| (\verb|NLanczosMode=2|), 
one can perform the first-step power-Lanczos calculations without (with) calculating correlation functions. 

\subsection{Models and Lattices}
We describe available Hamiltonians and lattices in mVMC.
General form of available Hamiltonians for mVMC is given by
{
\begin{align}
{\cal H}  &={\cal H}_T+{\cal H}_{I}, \\
{\cal H}_T&={\bf -}\sum_{i, j}\sum_{\sigma_i, \sigma_j}t_{ij\sigma_i\sigma_j} c_{i\sigma_i}^{\dag}c_{j\sigma_j},\\ 
{\cal H}_{\cal I}&=\sum_{i,j,k,l}\sum_{\sigma_1,\sigma_2, \sigma_3, \sigma_4}
{\cal I}_{ijkl\sigma_i\sigma_j\sigma_k\sigma_l}c_{i\sigma_i}^{\dagger}c_{j\sigma_j}c_{k\sigma_k}^{\dagger}c_{l\sigma_l},
\label{eq:GHam}
\end{align}
}
where ${c}_{i\sigma}^\dag$ (${c}_{i\sigma}$) is the creation (annihilation) 
operator of an electron on site $i$ with spin $\sigma= \uparrow$ or $\downarrow$.
This Hamiltonian includes the arbitrary one-body potentials
and two-body interactions in the particle-conserved systems.
General one-body potential $t_{ij\sigma_i \sigma_j}$ represents
the hopping between site $i$ with spin $\sigma_i$ and site $j$ with spin $\sigma_j$.
General two-body interaction 
{${\cal I}_{ijkl\sigma_i\sigma_j\sigma_k\sigma_l}$} represents
the interaction which annihilates a spin $\sigma_j$ 
\yoshimi{particle} at site $j$ and a spin $\sigma_l$ 
\yoshimi{particle} at site $l$, 
and creates a spin $\sigma_i$ particle at site 
$i$ and a spin $\sigma_k$ particle at site $k$.
Corresponding keywords for ${\cal H}_T$ and ${\cal H}_{\cal I}$
are shown in 
\tido{T}able ~\ref{table:Keywords}.
{In Standard mode, users can employ the anti-periodic boundary conditions and 
details are shown in \ref{sec:anti}.} 

{We note that localized spin-$1/2$ systems such as the Heisenberg model
can be regarded as a special case of the above Hamiltonian by completely
excluding the 
{doubly occupied}
sites at half filling with the Gutzwiller projections.
Therefore, by using the Gutzwiller projections,
one can use mVMC for solving spin-$1/2$ quantum spin models
by properly interpreting terms of {diagonal elements of}
$t_{ij\sigma_i\sigma_j}$ and ${\cal I}_{ijkl\sigma_i\sigma_j\sigma_k\sigma_l}$ 
as the potentials and interactions for localized spins\tido{, respectively}.}

Although
the Hamiltonian given in Eq.~(\ref{eq:GHam})
has the most general form, 
it is not efficient for specifying the simple interactions such as the
on-site Coulomb interactions by using the general form.
To reduce the efforts in making the input files,
we provide 
simple forms for the conventional interactions such
as 
the on-site Coulomb interactions (${\cal H}_U$),
the off-site Coulomb interactions (${\cal H}_V$),
the Ising-type Hund's rule couplings (${\cal H}_H$),
the exchange interactions (${\cal H}_E$), and
the pair hopping terms (${\cal H}_P$).
Forms of each interaction are
given as follows:
\begin{align}
{\cal H}_U   &= \sum_{i} U_i n_ {i \uparrow}n_{i \downarrow},\\
{\cal H}_V   &= \sum_{i,j} V_{ij}n_ {i}n_{j},\\
{\cal H}_H   &= {\bf -}\sum_{i,j}J_{ij}^{\rm Hund} (n_{i\uparrow}n_{j\uparrow}+n_{i\downarrow}n_{j\downarrow}),\\
{\cal H}_E   &= \sum_{i,j}J_{ij}^{\rm Ex} (c_ {i \uparrow}^{\dag}c_{j\uparrow}c_{j \downarrow}^{\dag}c_{i  \downarrow}+c_ {i \downarrow}^{\dag}c_{j\downarrow}c_{j \uparrow}^{\dag}c_{i  \uparrow}),\\
{\cal H}_P   &= \sum_{i,j}J_{ij}^{\rm Pair} 
(c_{i\uparrow}^{\dag}c_{j\uparrow}c_{i\downarrow}^{\dag}c_{j\downarrow}
+c_{j\downarrow}^{\dag}c_{i\downarrow}c_{j\uparrow}^{\dag}c_{i\uparrow})
,
\end{align}
where we define the charge density operator 
with spin $\sigma$ at site $i$ as $n_{i \sigma}=c_{i\sigma}^{\dag}c_{i\sigma}$ and 
the total charge density operator at site $i$ as $n_i=n_{i\uparrow}+n_{i\downarrow}$. 
Corresponding keywords are shown in
\tido{T}able \ref{table:Keywords}.

In Standard mode,
users can select the standard models 
with the human-readable {keywords}, i.e.,
Hubbard-type models 
\tido{are} specified with \verb|model="Hubbard"|,
localized spin-$1/2$ Heisenberg models 
\tido{are} specified with \verb|model="Spin"|, and
Kondo-lattice models 
\tido{are} specified with \verb|model="Kondo"|.
By adding \verb|GC| at the back of the {keywords} for models, for example \verb|model="HubbardGC"|,
one can treat the $S^{z}$ non-conserved systems.
We note that \verb|GC| is an abbreviation grand canonical ensemble
but mVMC only supports the  $S^{z}$ non-conserved systems and
does not support the particle non-conserved systems in ver.1.0.
We summarize available {keywords} for lattices and models 
in Standard mode in 
\tido{T}able.\ref{table:model} . 

\begin{table}[tb!]
\centering
\caption{Examples of key\tido{word}s for models and lattices in Standard mode.
~GC means $S^{z}$ non-conserved system.}
\label{my-label}
\begin{tabular}{ll}
\hline \hline
types & keywords \\
\hline \hline
model/canonical         & Hubbard, Spin, Kondo \\
\hline
model/grand canonical  & HubbardGC, SpinGC, KondoGC   \\
\hline
\multirow{2}{*}{lattice}
                & chain, square, triangular, \\
                & honeycomb, kagome \\
\hline \hline
\end{tabular}
\label{table:model}
\end{table}

The form of the Hamiltonians with ``Hubbard/HubbardGC" 
is given by
\begin{align}
&{\cal H}_{\rm Hubbard}= -\mu \sum_{i \sigma} \ {c}^\dagger_{i \sigma} \ {c}_{i \sigma}-\sum_{ij, \sigma} t_{i j} \ {c}^\dagger_{i \sigma} \ {c}_{j \sigma} \\ \notag
&+ \sum_{i} U \ {n}_{i \uparrow} \ {n}_{i \downarrow} + \sum_{ij} V_{i j} \ {n}_{i} \ {n}_{j}, \\
\end{align}
where $V_{ij}$ denotes the off-site Coulomb interactions
and its range 
depend\tido{s} on the lattice structures.
The form of the Hamiltonians with ``Spin/SpinGC" 
is given by
\begin{align}
&\ {\cal H}_{\rm Spin} = -h \sum_{i} \ {S}_{i}^{z} - \Gamma \sum_{i} \ {S}_{i}^{x} 
+ D \sum_{i} \ {S}_{i}^{z} \ {S}_{i}^{z} \\ \notag
&+ \sum_{ij, \alpha}J_{i j \alpha} \ {S}_{i}^{\alpha} \ {S}_{j}^{ \alpha}+ \sum_{ij, \alpha \neq \beta} J_{i j \alpha \beta} \ {S}_{i}^{ \alpha} \ {S}_{j}^{ \beta}.
\end{align}
Here, spin operators are defined as follows:
\begin{align}
&S_{i}^{x}=(c_{i\uparrow}^{\dagger}c_{i\downarrow}+c_{i\downarrow}^{\dagger}c_{i\uparrow})/2,\\
&S_{i}^{y}={\rm i}(-c_{i\uparrow}^{\dagger}c_{i\downarrow}+c_{i\downarrow}^{\dagger}c_{i\uparrow})/2,\\
&S_{i}^{z}=(n_{i\uparrow}-n_{i\downarrow})/2,\\
&S_{i}^{+}=S_{i}^x+{\rm i}S_{i}^{y},\\
&S_{i}^{-}=S_{i}^x-{\rm i}S_{i}^{y}.
\end{align}
The form of the Hamiltonians with ``Kondo/KondoGC" 
is given by
\begin{align}
&\ {\cal H}_{\rm Kondo} = - \mu \sum_{i \sigma} \ {c}^\dagger_{i \sigma} \ {c}_{i \sigma}-\sum_{i j,\sigma} t_{ij}\ {c}^\dagger_{i \sigma}\ {c}_{j \sigma} \\ \notag
&+ \sum_{i} U \ {n}_{i \uparrow} \ {n}_{i \downarrow}
+ \sum_{i j} V_{i j} \ {n}_{i} \ {n}_{j} \\ \notag
&+ \frac{J}{2} \sum_{i} \left\{\ {S}_{i}^{+} \ {c}_{i \downarrow}^\dagger \ {c}_{i \uparrow}+\ {S}_{i}^{-} \ {c}_{i \uparrow}^\dagger \ {c}_{i \downarrow}
+\ {S}_{i}^{z} (\ {n}_{i \uparrow} - \ {n}_{i \downarrow})\right\},\\
\end{align}
where spin operators for the localized spins are
denoted by $S^{\alpha}$ and
the operators for itinerant electrons are
denoted by $c_{i}^{\dagger}/c_{i}$ and $n_{i}$.
Details of the parameters are given in manuals~\cite{ma}.

\subsection{Visualization tools in mVMC}

\subsubsection{Display lattice {geometry}}
By using \verb|lattice.gp|, which is generate\tido{d} after executing \verb|vmc.out| or \verb|vmcdry.out|,
we can display the 
{geometry} of the simulation cell generated by Standard mode as follows:
\begin{verbatim}
$ gnuplot lattice.gp
\end{verbatim}
{Here, we use gnuplot~\cite{gnuplot} for visualization.}
Then \tido{the} lattice 
{geometry} is displayed in the new window as shown in Fig.~\ref{fig:lattice_gp}.
Figure \ref{fig:lattice_gp} shows the lattice 
{geometry} of the 20-sites 
Hubbard model on the square lattice and we use the following input file:
\begin{verbatim}
model = "Hubbard"
lattice = "square"
a0w = 4
a0l = 2
a1w = -2
a1l = 4
t = 1.0
U = 10.0
nelec = 20
\end{verbatim}

\begin{figure}[tb!]
  \begin{center}
    \includegraphics[width=6cm]{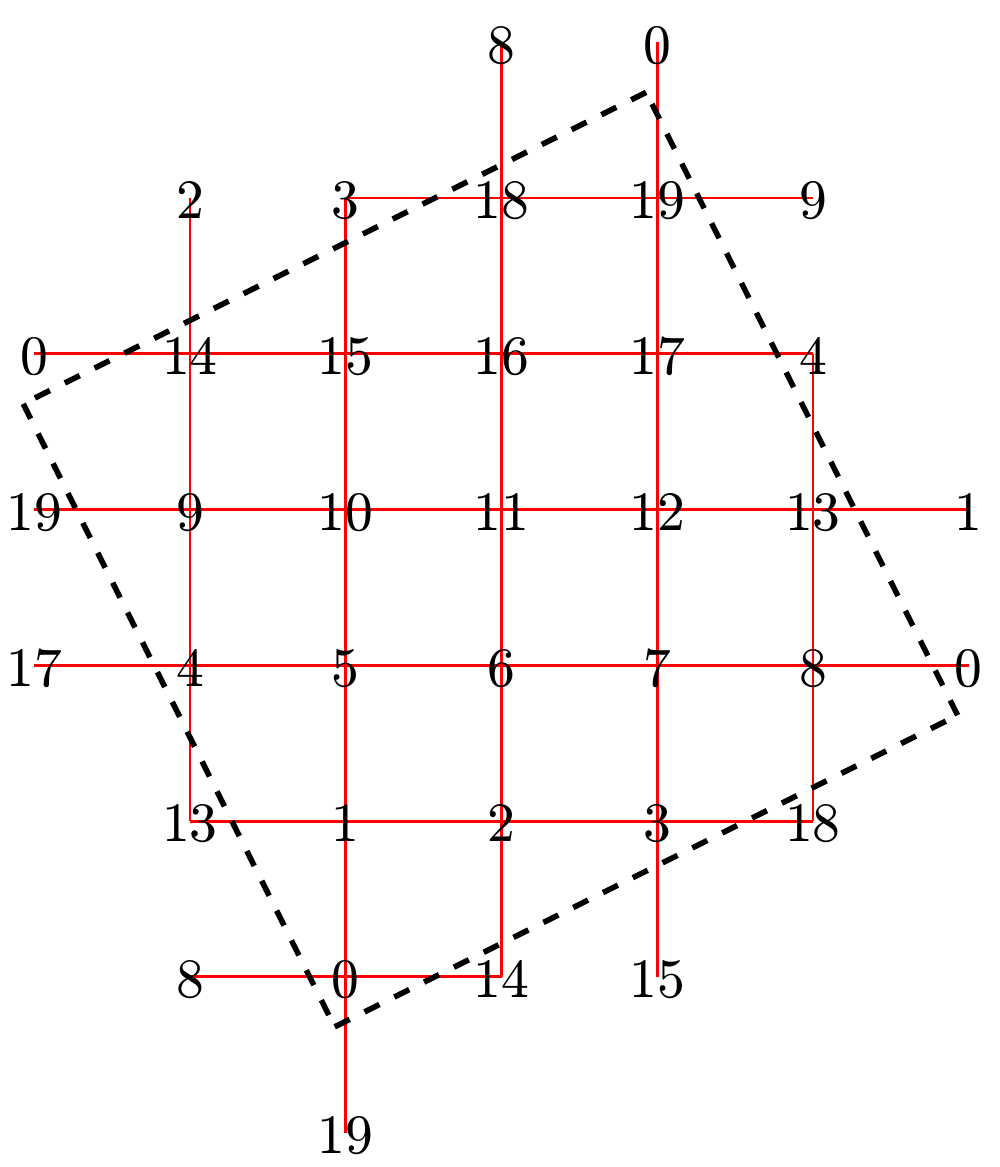}
    \caption{
      {Geometry} of the simulation cell on the 20-site square lattice.
      The red solid line and the black dashed line indicate the nearest-neighbor hopping and
      the boundary of the simulation cell, respectively.}
    \label{fig:lattice_gp}
  \end{center}
\end{figure}

\subsubsection{Fourier transformation of correlation functions}

\begin{figure}[tb!]
  \begin{center}
    \includegraphics[width=7.5cm]{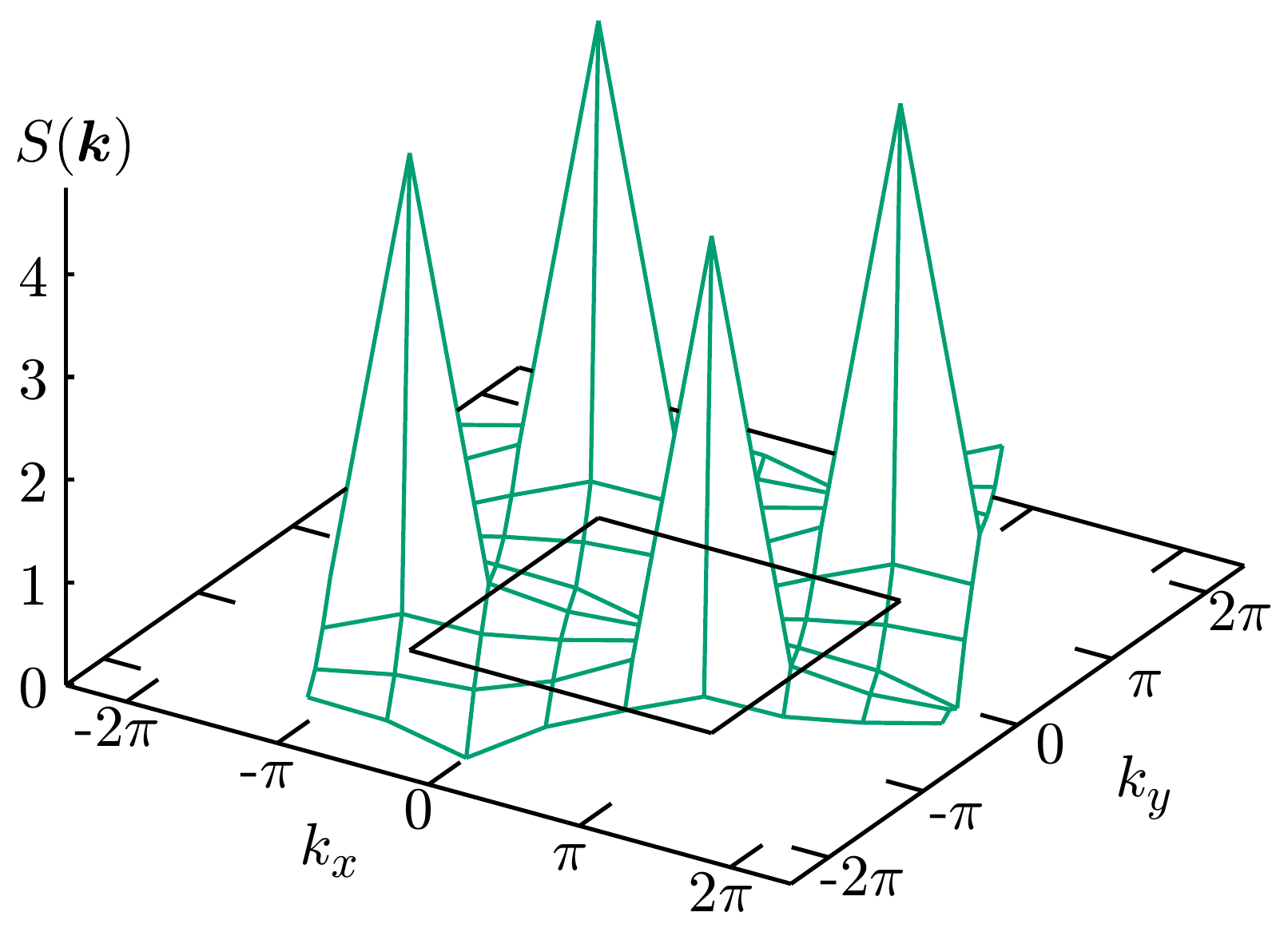}
    \caption{
    The spin-spin correlation function of the {20-site} Hubbard model on the square lattice.
    {The (black) solid line at the bottom} indicates the {first} Brillouin-zone boundary.}
    \label{fig:sksk}
  \end{center}
\end{figure}

mVMC has utility programs (fourier tool) to 
compute the structure factors in the reciprocal space, which are defined as
\begin{align}
  X(\vec{k})
  \equiv \frac{1}{N_{\rm cell}} \sum_{i,j}^{N_{\rm site}} e^{-{\rm i} {\vec{k}}\cdot({\vec{ R}}_i - {\vec{R}}_j)}
  \langle { X}_{i}^\dagger {X}_{j}\rangle,
\end{align}
where 
\tmorita{${X_i}$} is an operator such as ${c}_{i\uparrow}$ and  ${c}_{i\uparrow}^{\dagger}{c}_{i\uparrow}$.
{We note that $N_{\rm cell}$ denotes {the} number of the unit cells, for example,
$N_{\rm site}=2\times N_{\rm cell}$ for honeycomb lattice.} 
We can compute this type of 
correlation function by using fourier tool.
These programs support the calculation\tido{s} of 
the one-body correlation (momentum distribution) $n_{\sigma}(\vec{k})$,
the charge structure factors $N(\vec{k})$, and
the spin structure factors $S(\vec{k})$, $S^{xy}(\vec{k})$, $S^{z}(\vec{k})$,
which are defined as 
\begin{align}
n_{\sigma}(\vec{k})&= \frac{1}{N_{\rm cell}} \sum_{i,j}^{N_{\rm site}} e^{-{\rm i} \vec{k}\cdot(\vec{R}_i - \vec{R}_j)}\langle {c}_{i\sigma}^\dagger {c}_{j\sigma}\rangle,\\
N(\vec{k})&= \frac{1}{N_{\rm cell}} \sum_{i,j}^{N_{\rm site}} e^{-{\rm i} \vec{k}\cdot(\vec{R}_i - \vec{R}_j)}\langle {n}_{i\sigma}^\dagger {n}_{j\sigma}\rangle,\\
S(\vec{k})&= \frac{1}{N_{\rm cell}} \sum_{i,j}^{N_{\rm site}} e^{-{\rm i} \vec{k}\cdot(\vec{R}_i - \vec{R}_j)}
     \langle \vec{S}_{i}\cdot\vec{S}_{j}\rangle,\label{eq:Sk}\\
S^{xy}(\vec{k})&= \frac{1}{N_{\rm cell}} \sum_{i,j}^{N_{\rm site}} e^{-{\rm i} \vec{k}\cdot(\vec{R}_i - \vec{R}_j)}
     \langle{S}^{+}_{i}S^{-}_{j}+{S}^{-}_{i}S^{+}_{j}\rangle,\\
S^{z}(\vec{k})&= \frac{1}{N_{\rm cell}} \sum_{i,j}^{N_{\rm site}} e^{-{\rm i} \vec{k}\cdot(\vec{R}_i - \vec{R}_j)}
     \langle {S}^{z}_{i}{S}^{z}_{j}\rangle.
\end{align}

We demonstrate an example of {calculation for} $S(\vec{k})$
{on} the 
\tmorita{20-site} Hubbard model on the square lattice shown in the previous section.
After 
{calculating}
the correlation function in the real space by using \verb|vmc.out|
with \verb|NVMCCalMode=1|,
the utility program \verb|fourier| is executed in the same directory as follows:
\begin{verbatim}
$ fourier namelist.def geometry.dat
\end{verbatim}
where \verb|geometry.dat| is the file specifying the positions of sites, which
is generated automatically in Standard mode.
Then the fourier-transformed correlation function 
is stored in a file \verb|output/zvo_corr.dat|,
where \verb|zvo| is the prefix specified in the input file.
For two dimensional systems,
we can display a color-plot by using another utility program \verb|corplot| as follows:
\begin{verbatim}
$ corplot output/zvo_corr.dat
\end{verbatim}
In this program, we cho\tido{o}se the type of 
correlation function listed above,
and can draw the correlation functions  as show\tido{n} in Fig.~\ref{fig:sksk}.

\section{Basics of mVMC}
\subsection{Sampling method}
\subsubsection{Monte Carlo sampling}
In the VMC method, the importance sampling based on the Markov-chain Monte Carlo is used for evaluating the 
expectation values for the many-body wave functions~\cite{Gros_AP1989,Toulouse_2015}.
As a complete basis, we use the real-space configuration 
$\{| x\rangle\}$ defined as 
\begin{equation}
|x\rangle =  \prod_{n,\sigma} c_{\vec{r}_{n\sigma}}^{\dag} \ket{0},
\end{equation}
where $\tido{\vec{r}}_{n\sigma}$ denotes the position of the $n$th electron with spin $\sigma$.
For the $S^{z}$-conserved system, 
we used the fixed $S^{z}$ real-space configuration defined as
\begin{equation}
  |x\rangle =  \prod_{n=1}^{N_{\tido{\rm up}}} c_{\vec{r}_{n\uparrow}}^{\dag} \prod_{n=1}^{N_{\tido{\rm down}}} c_{\vec{r}_{n\downarrow}}^{\dag} \ket{0},
\end{equation}
where $N_{\rm up}$ ($N_{\rm down}$) denotes the number of 
up (down) electrons and {the} total {value of} $S^{z}$ is given by $S^{z}=(N_{\rm up}-N_{\rm down})/2$.
By using $\{| x\rangle\}$, 
we rewrite the expected value of the operator $A$ as follows:
\begin{align}
&\langle A \rangle =\frac{\langle \psi| A| \psi \rangle}{\langle \psi | \psi \rangle} 
=\sum_x \frac{\langle \psi| A | x\rangle \langle x| \psi \rangle}{\langle \psi |\psi \rangle} 
=\sum_x \rho(x) \frac{\langle \psi| A | x\rangle }{\langle \psi |x \rangle},\\
&\rho(x)=\frac{|\braket{x|\psi}|^2}{\braket{\psi|\psi}}.
\end{align}
By performing the Markov-chain Monte Carlo with respect 
to the weight $\rho(x)$, i.e., by generating the real-space configuration\tmorita{s}
according to the weight $\rho(x)$,
we can evaluate $\braket{A}$ as 
\begin{equation}
\braket{A}\sim\frac{1}{N_{\rm MC}}\sum_{x}\frac{\langle \psi| A | x\rangle }{\langle \psi |x \rangle},
\label{Eq:MC}
\end{equation}
where $N_{\rm MC}$ is the number of Monte Carlo samplings.
In mVMC, we use the Mersenne twister~\cite{Mutsuo_SFMT2008} for 
generating the pseudo random numbers.

\subsubsection{Update of real-space configurations}
\begin{figure}[tb!]
  \begin{center}
    \includegraphics[width=7.5cm]{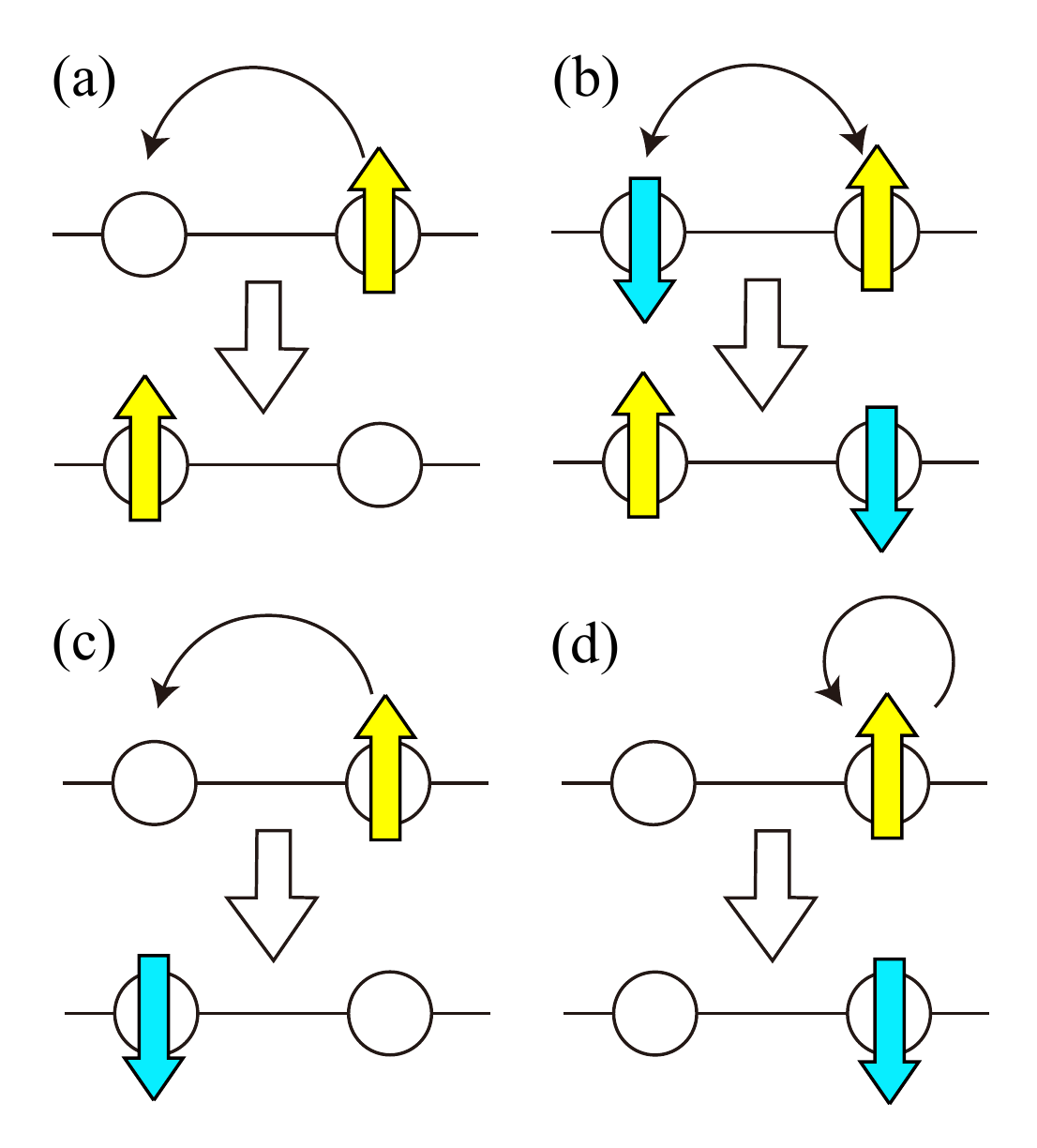}
    \caption{
(a)~Hopping update.
(b)~Exchange update. 
(c)~Hopping update with spin flip.
(d)~Local spin flip update.}
    \label{fig:update}
  \end{center}
\end{figure}

{For the itinerant electron systems such as the Hubbard model,
we update the real-space configurations $\ket{x}$ with the hopping process, i.e.,
one electron hops into another site as shown in Fig.~\ref{fig:update}(a).
In addition to the hopping update,
we can use the exchange update, i.e.,
opposite spins exchange as shown in Fig.~\ref{fig:update}(b).
For the local spin models such as the 
Heisenberg model, \tido{the} hopping update
is prohibited and only the
exchange update is allowed.
Even in the itinerant electrons models
\tido{with strong on-site interaction},
it is necessary to use the exchange update
for an efficient Monte Carlo sampling because the creation
of the doubly occupied site (or equivalently the creation of the holon site) 
becomes {a} rare event in the strong coupling region.}

In the $S_{z}$ non-conserved system\tido{s},
we employ the hopping update with spin flip 
as shown in Fig.~\ref{fig:update} (c).
We also employ local spin flip update 
\tmorita{as shown} in Fig.~\ref{fig:update}(d).
Although the local spin flip is included in the 
hopping with spin flip, 
for an efficient sampling, 
it is necessary to  explicitly perform the local spin flip.

As we show later, 
the inner product between the Pfaffian wave functions
and the real-space configuration $\ket{x}$ is given by the Pfaffian
of the skew symmetric matrix $X$.
It is time consuming to calculate the 
Pfaffian for each real-space configuration.
Because the changes in the real-space configuration
induce the changes {only} in a few rows and columns in $X$,
{numerical cost can be reduced
by using {the} Sherman-Morrison-type update technique.}
Details of the fast update techniques of the 
Pfaffian wave functions are shown in 
\tido{R}efs.~\cite{Tahara_JPSJ2008,Morita_JPSJ2015}.

\subsection{Wave functions}
In mVMC, the form of the variational wave function is given as
\begin{align}
&|\psi \rangle = {\cal P}{\cal L} |\phi_{\rm Pf} \rangle,
\end{align}
where 
$|\phi_{\rm Pf} \rangle$ denotes the
{pair-product} part of the wave functions,
${\cal L}$ denotes the quantum-number projectors
such as the total-spin and momentum projections, 
and
${\cal P}$ denotes the correlation factors such as the Gutzwiller and the 
Jastrow factors.
We detail each part of the wave functions in the following.

\subsubsection{{Pair-product} part -- Pfaffian wave functions}
The {pair-product} part of the wavefunction in mVMC is 
represented as the Pfaffian wave function, which is defined as
\begin{equation}
|\phi_{\rm Pf} \rangle = \left[\sum_{i, j=0}^{\Ns-1} 
\sum_{\sigma_i, \sigma_j}F_{i\sigma_ij \sigma_j}c_{i\sigma_i}^{\dag}c_{j\sigma_j}^{\dag} \right]^{\Ne/2}|0 \rangle,
\label{Eq:Pf}
\end{equation}
where $\Ne$ is the number of electrons and
$\Ns$ is the number of sites.
Index $i$ denotes the number of sites and
$\sigma_{i}$ denotes the spin at $i$th site.
For simplicity, we denote the 
combination of the site index and 
spin index by the capital letter such as $I=(i,\sigma_{i})$ or
$I=i+\Ns\times\sigma_{i}~(0\leq i\leq \Ns-1,\sigma_{i}=0,1)$ in the following.
We note that site index $i$ can include 
the orbital degrees of freedom 
in the multi-orbital systems.
If \yoshimi{the} total \yoshimi{value of} $S^{z}$ is conserved and fixed to 0, 
we often use the {anti-parallel} Pfaffian wave function defined as
\begin{equation}
|\phi_{\rm AP-Pf} \rangle = \left[\sum_{i, j=0}^{\Ns-1} 
f_{ij}c_{i\uparrow}^{\dag}c_{j\downarrow}^{\dag} \right]^{\Ne/2}|0 \rangle.
\label{Eq:AP-Pf}
\end{equation}

The Pfaffian wave function is an extension of the 
Slater determinant defined as 
\begin{align}
|\phi_{\rm SL}\rangle&=\Big(\prod_{n=1}^{\Ne}
\psi_{n}^{\dagger}\Big)|0\rangle,~~\psi_{n}^{\dagger}
=\sum_{I=0}^{2\Ns-1}\Phi_{In}c^{\dagger}_{I},
\end{align}
where $I$ denotes the {index} including the site and spin {index}
and $n$ denotes the index of the occupied orbital.
As we show in \ref{sec:PfAndSlater}, 
from the Slater determinant, 
it is possible to construct the equivalent
Pfaffian wave function by using the relation\tido{s}
\begin{align}
&F_{IJ}=\sum_{n=1}^{{\Ne}/{2}}\Big(\Phi_{I,2n-1}\Phi_{J,2n}-\Phi_{J,2n-1}\Phi_{I,2n}\Big),\\
&f_{ij}=\sum_{n=1}^{N_{\tido{\rm e}}/2}\Phi_{in\uparrow}\Phi_{jn\downarrow}.
\label{eq:fijSL}
\end{align}
This result shows that the Pfaffian wave functions include the 
Slater determinants.
{We note that the Pfaffian wave functions can describe states that are not 
described by the Slater determinant such as the superconducting 
phases with fixed particle numbers.}

Inner product between $\Ne$-particle 
real-space configuration $|x\rangle$ and 
the Pfaffian wave functions \imada{is} 
described by the Pfaffian as follows:
\begin{align}
\langle x|\phi_{\rm Pf}\rangle = (\Ne/2)!{\rm Pf}(X),
\end{align}
where $X$ is a $2\Ne\times 2\Ne$ skew-symmetric matrix $X$
whose  elements 
\tido{are} given as
\begin{align}
X_{IJ}=F_{IJ}-F_{JI}.
\end{align}
This relation is the
reason why $\ket{\phi_{\rm Pf}}$ is called \tido{the} Pfaffian wave function.
We note that the Pfaffian is defined for the skew-symmetric matrices and
its square is the determinant of the skew-symmetric matrices, i.e.,
\begin{align}
{[{\rm Pf}(X)]^{2}}={\rm det}(X).
\end{align}
We use 
Pfapack~\cite{Wimmer_PFAPACK2012} for calculating Pfaffian\tido{s} in mVMC.
In the field of the quantum chemistry,
the Pfaffian wave 
functions is often called geminal wave functions~\cite{Sorella_JCS2007,Bajdich_PRB2008}.

Although the Pfaffian wave function has $2N_{s}\times(N_{s}-1)$ 
complex variational parameters,
to reduce the numerical cost,
it is possible to impose the sublattice structures
\tido{on} the Pfaffian wave functions.
For example, 
if we consider the $2\times2$ sublattice structures 
{in the real space} for the
$N_{\rm s}=L\times L$ square lattice,
the number of the variational parameters are reduced from
$O(N_{\rm s}^2)$ to $O(N_{\rm s})$.
In Standard mode, 
by specifying the {keywords} $W_{\rm sub}$ ($L_{\rm sub}$),
one can impose the sublattice structures 
{in $x$ direction ($y$ direction) for the variational parameters}.
We note that the sublattice structure allows
the orders within the sublattice structures, i.e.,
in $2\times 2$ sublattice {in the real space}, the 
{ordering} wave vectors {in the momentum space} are limited
to $(\pi,\pi)$, $(0,\pi),(\pi,0)$, and $(0,0)$.
If one wants to examine stability of the long-period orders,
it is better to take larger sublattice structures or not
to impose the sublattice structure if possible.

\subsubsection{Quantum number projection}
The quantum-number projector ${\cal L}$ consists of
the momentum projector ${\cal L}_K$,
the point-group symmetry projector ${\cal L}_P$, and
the total-spin projector ${\cal L}_S$, which are defined as
\begin{align}
&{\cal L}={\cal L}_{S}{\cal L}_K{\cal L}_P,\\
&{\cal L}_K=\frac{1}{\Ns}\sum_{{\bm R}}e^{{\rm i}\vec{K}\cdot\vec{R}}{T}_{\vec{R}},\label{Eq:momP}\\
&{\cal L}_P=\frac{1}{N_{g}}\sum_{\vec{p}}g_{\vec{p}}(\alpha)^{-1}T_{\vec{p}},\label{Eq:pointP}\\
&{\cal L}_S=\frac{2S+1}{8 \pi^2}\int d\Omega {P_{S}}(\cos \beta) {R}(\Omega)\label{Eq:spnP}.
\end{align}
Here, {$\vec{K}$} is the total momentum 
of the whole system and 
$\tido{T}_{\bm R}$ is the translational 
operator corresponding to the translational vector ${\bm R}$, 
$T_{\vec{p}}$ is the translational operator 
{corresponding to the vector point-group operations $\vec{p}$,}
and $g_{\vec{p}}$ is the character for point-group operations.
The number of 
elements in the
point group or translational group is denoted by $N_{g}$.
We only consider the total-spin projection which filters out {the} 
$S^{z}=0$ component in the wave function and generate the wave function
with $S^{z}=0$ and the total spin $S$.
In general, the total-spin projection which filters out
$S^{z}=M$ component and generate the wave function
with $S^{z}=M^{\prime}$ and {the total spin} $S$ is possible. Details of 
such general spin projections are shown in the literature~\cite{RingSchuck,Mizusaki_PRB2004}.
In the definition of ${\cal L}_{S}$,
$\Omega=(\alpha, \beta, \gamma)$ denotes the Euler angles, 
${R}(\Omega)=e^{{\rm i}\alpha S^{z}}e^{{\rm i}\beta S^{y}}e^{{\rm i}\gamma S^{z}}$ is 
the rotational operator {in the spin space}, 
$P_S(x)$ is the $S$th Legendre polynomial, respectively.

Here, we explain the essence of the quantum-number projections by taking 
{a} point-group projections as an example.
{
When the Hamiltonian preserves a symmetry, any exact eigenstate 
has to respect its quantum number associated 
with the symmetry, while wavefunctions constructed so far do 
not necessarily satisfy this requirement. 
Such a symmetry-preserved 
state with a given quantum number can be constructed from a given state $|\phi\rangle$}
as
\begin{align}
|\phi\rangle=\sum_{\alpha}a_{\alpha}|\alpha\rangle,~T_{\vec{p}} |\alpha\rangle=g_{\alpha}({\vec{p}}) |\alpha\rangle,
\end{align}
where $\ket{\alpha}$ 
{are}
the eigenvectors 
 {of} $T_{\vec{p}}$ and
$g_{\alpha}({\vec{p}})$ 
{are}
the eigenvalues.
Here, $T_{\vec{p}}$ is defined as
\begin{align}
T_{\vec{p}}c_{\vec{r}}^{\dagger}T_{\vec{p}}^{-1}=c_{\vec{r}+\vec{p}}^{\dagger}.
\end{align}

To extract  $|\alpha\rangle$ from $|\phi\rangle$,
we define the projection operator as
\begin{align}
{\cal L}_{\alpha}=\frac{1}{N_g}\sum_{\vec{p}}g_{\alpha}(\vec{p})^{-1}T_{\vec{p}},
\end{align}
where $N_g$ is the number of 
elements in the
point group or translational group.
By using the projection operator,
we can show the following relation.
\begin{align}
{\cal L}_{\alpha}\ket{\phi}=
\frac{1}{N_g}\sum_{\vec{p},\alpha^{\prime}}
g_{\alpha}(\vec{p})^{-1} g_{\alpha^{\prime}}(\vec{p})a_{\alpha^{\prime}}\ket{\alpha^{\prime}} 
= a_{\alpha}\ket{\alpha}.
\end{align}
Here, we use the orthogonal relation for characters~\cite{Dresselhaus_2007}
\begin{align}
\sum_{\vec{p}}g_{\alpha}(\vec{p})^{-1} g_{\alpha^{\prime}}(\vec{p})=N_g\delta_{\alpha,\alpha^{\prime}}.
\end{align}

For the momentum projection, 
we take {the} translational operators $T_{\vec{R}}$ defined as
\begin{align}
T_{\vec{R}}c_{\vec{r}}^{\dagger}T_{\vec{R}}^{-1}=c_{\vec{r}+\vec{R}}^{\dagger},
\end{align}
where $\vec{R}$ denotes the translational vector and
{the} corresponding character is $e^{{\rm i}\vec{k}\tido{\cdot}\vec{R}}$.
By using  $T_{\vec{R}}$ and $e^{{\rm i}\vec{k}\tido{\cdot}\vec{R}}$, 
the momentum projection is defined by Eq.~(\ref{Eq:momP}).

To define the momentum and point-group projections,
it is necessary to specify
the translational operations ($T_{\vec{p}}$) and the associate{d}
weight $g_{\alpha}(\vec{p})$
in the \verb|qptrans.def| with {the} keyword \verb|TransSym|.
By preparing the {file} \verb|qptrans.def|,
one can perform the desired 
momentum and the point-group projections. 
{For the anti-periodic conditions, it is necessary to consider the changes of signs in
the Pfaffian wave functions. Details are {shown} in \ref{sec:anti}.}

For the total-spin projection, 
sum of the discrete freedoms becomes the 
integration \imada{over} 
the Euler angles in the spin space and
translational operation becomes the rotational operator $R(\Omega)$
in the spin space.
 The associated character is given by \tohgoe{$P_{S}(\cos{\beta})$}.
Thus, the spin projection is given in Eq.~(\ref{Eq:spnP}).
Although it is possible to perform the spin projection
to the general Pfaffian wave function defined in Eq.~(\ref{Eq:Pf}),
where $S^{z}$ component is not conserved,
mVMC ver. 1.0 only supports spin projection for 
anti-parallel  Pfaffian wave functions defined in Eq.~(\ref{Eq:AP-Pf}).
Because total $S^{z}$ in $|\phi_{\rm A-Pf} \rangle$ is  
{definitely} zero,
triple integration 
{for} the spin projection 
{is reduced to a}
single integration as follows:
\begin{align}
&{\cal L}_S|\phi_{\rm A-Pf} \rangle =\sum_{x}\ket{x}\frac{2S+1}{2}\int_{0}^{\pi} d\beta P_s(\cos \beta)\bra{x}e^{i\beta S^{y}}\ket{\phi_{\rm A-Pf}}.
\end{align}
{The integration is performed by using the Gauss-Legendre formula.}
We note that anti-parallel Pfaffian wave function is transformed as
\begin{align}
e^{i\beta S^{y}}\ket{\phi_{\rm A-Pf}}=
\left[\sum_{i, j} 
\sum_{\sigma_i, \sigma_j}f_{ij}
\times\kappa(\sigma_{i},\sigma_{j})c_{i\sigma_{i}}^{\dag}c_{j\sigma_{j}}^{\dag} \right]^{\Ne/2}\ket{0},
\end{align}
where $\kappa(\sigma_{i},\sigma_{j})$ is defined as
\begin{align}
\kappa(\uparrow,\uparrow)&= -\cos{(\beta/2)}\sin{(\beta/2)},\\
\kappa(\uparrow,\downarrow)&= \cos{(\beta/2)}\cos{(\beta/2)},\\
\kappa(\downarrow,\uparrow)&= \sin{(\beta/2)}\sin{(\beta/2)},\\
\kappa(\downarrow,\downarrow)&= \cos{(\beta/2)}\sin{(\beta/2)}.
\end{align}

\subsubsection{Correlation factors}
To take into account the many-body correlations,
the Gutzwiller factors ${\cal P}_G$~\cite{Gutzwiller_PRL1963}, 
the Jastrow factors ${\cal P}_J$~\cite{Jastrow_PR1955,PRL_Capello2005},
the $m$-site doublon-holon correlation factors ${\cal P}_{d-h}^{(m)}$ ($m=2,4$)~\cite{Yokoyama_JPSJ1990}
are implemented in mVMC~\cite{Tahara_JPSJ2008}, which are defined as follows:
\begin{align}
&{\cal P}={\cal P}_G{\cal P}_J{\cal P}_{d-h}^{(2)}{\cal P}_{d-h}^{(4)},\\
&{\cal P}_G=\exp\left[ \sum_i g_i n_{i\uparrow} n_{i\downarrow} \right],\\
&{\cal P}_J=\exp\left[\frac{1}{2} \sum_{i\neq j} v_{ij} (n_i-1)(n_j-1)\right],\\
&{\cal P}_{d-h}^{(m)}= \exp \left[ \sum_t \sum_{n=0}^m (\alpha_{mnt}^d \sum_{i}\xi_{imnt}^d+\alpha_{mnt}^h \sum_{i}\xi_{imnt}^h)\right].
\end{align}
{In the definitions of the doublon-holon correlation factors,
$\alpha_{mnt}^{d}$ and $\alpha_{mnt}^{h}$, $\alpha_{4nt}^{d}$ are the variational parameters.
Real-space diagonal operators $\xi_{imnt}^{d}$ and $\xi_{imnt}^{h}$
are defined as follows:
When a doublon (holon) exists at the $i$th site
and $n$ holons (doublons) exist at the $m$-site neighbors defined by $t$ around $i$th site,
$\xi_{imnt}^{d}$ ($\xi_{imnt}^{h}$) become 1.
Otherwise $\xi_{imnt}^{d}$ or $\xi_{imnt}^{h}$ are 0.
}
We note {that} the above correlation factors are 
diagonal 
{in}
real-space 
{representation}, i.e.,
\begin{align}
{\cal P}\ket{x}=P(x)\ket{x},
\end{align}
where the $P(x)$ is 
 {a} 
\tohgoe{scalar} number.

By taking $g_{i}=-\infty$, it is possible 
to completely 
{eliminate} 
the doublons in the real-space configurations.
Using this projection, we describe the spin 1/2 localized spins. 
In mVMC,
one can specify the positions of the local spins
in \verb|LocSpn.def| with the keyword \verb|LocSpn|.

It has been shown that
the long-range Jastrow factors play important roles in describing
the Mott insulating phase~\cite{PRL_Capello2005}.
If the Jastrow factor becomes {long ranged}, i.e.,
the indices of the Jastrow factor run {over} 
all the sites,
$v_{ij}$ often {gets} large and induce numerical instability.
\imada{By subtracting}
constant value from $v_{ij}$ and $g_{i}$,
\imada{we are able to stabilize the numerical computation without modifying the results.}

\subsection{Optimization method}
In this subsection, we describe the optimization
method used in mVMC. 
The SR method 
proposed by 
{Sorella}~\tido{\cite{Sorella_PRB2001,Sorella_JCS2007}} 
\tido{is} an efficient way for optimizing many variational parameters.
By using the SR method, it has been shown that
more than ten thousands variational parameters 
can be simultaneously optimized. 
The SR method largely relaxes the restrictions in the 
variational wave functions, which 
{allows 
{to improve}} 
the accuracy 
of the variational Monte Carlo method.

The essence of the SR method is the imaginary-time evolution of the
wave functions in the restricted Hilbert space that is spanned by the 
variational parameters as we explain in sec.~\ref{sec:SR}.
This means that the effective dimensions of the Hilbert space increase 
by increasing the number of the variational parameters.
Thus, in principle, the accuracy of the {imaginary-time} evolution becomes high for
the wavefunctions with many-variational parameters.
In the SR method, 
because 
{taking derivatives}
of wavefunctions
is an important procedure,
we detail how to differentiate the wave functions in sec.~\ref{sec:SR}
and \ref{sec:diffPf}.

In the SR method, 
it is necessary to solve the linear algebraic equation with respect to
the overlap matrix $S$.
Because the \tohgoe{dimension} of $S$ \tohgoe{is}
square of {the} \tohgoe{number of} the variational parameters,
storing the matrix $S$ is the most \tohgoe{memory-consuming} part in mVMC and 
it determines the applicable range of mVMC.
To relax the limitation, 
a conjugate gradient (CG) method for the SR method is proposed (SR-CG method)~\cite{Neuscamman_PRB2012}.
In this method, because it is not necessary to explicitly 
{compute} $S$,
the required memory in mVMC is dramatically reduced.
By using this method, it is now possible to
optimize the variational parameters up to the order of hundred thousands.
We detail the SR-CG method in sec.~\ref{sec:CG}.


\subsubsection{Stochastic reconfiguration method}
\label{sec:SR}
The imaginary-time-dependent Schr\"{o}dinger equation
is given by
\begin{align}
\frac{d}{d\tau}\ket{{\psi}(\tau)}=-{H}\ket{{\psi}(\tau)}.
\end{align}
By substituting the normalized wave
function $\ket{\bar{\psi}(\tau)}$ defined by
\begin{align}
\ket{\bar{\psi}(\tau)}=\frac{\ket{\psi(\tau)}}{\sqrt{\braket{\psi(\tau)|\psi(\tau)}}}
\end{align}
into the Schr\"{o}dinger equation, we obtain
\begin{align}
\frac{d}{d\tau}\ket{\bar{\psi}(\tau)}=-({H}-\langle{H}\rangle)\ket{\bar{\psi}(\tau)}.
\end{align}
If the wavefunction $\ket{\bar{\psi}(\tau)}$ is represented by 
the variational parameter $\alpha(\tau)$, 
the Schr\"{o}dinger equation is rewritten as 
\begin{align}
\sum_{k}\dot{\alpha_k}\frac{\partial }{\partial \alpha_{k}}\ket{\bar{\psi}(\tau)}=-({H}-\langle{H}\rangle)\ket{\bar{\psi}(\tau)}.
\end{align}
where $\dot{\alpha_k}$ is the derivative of the $k$th variational parameter $\alpha_k$ with respect to $\tau$.

By minimizing the $L_{2}$-norm of the Schr\"{o}dinger equation with respect to $\dot{\alpha}_{k}$, i.e.,
\begin{align}
\min_{\dot{\alpha}_{k}}\left\|\sum_{k}\dot{\alpha}_k\ket{\partial_{\alpha_k}\bar{\psi}(\alpha(\tau))}+({H}-\langle{H}\rangle)\ket{\bar{\psi}(\alpha(\tau))}\right\|,
\end{align} 
we can obtain the best imaginary-time evolution in the restricted Hilbert space.
This minimization principle is called time-dependent variational 
principle (TDVP)~\cite{Mclachlan_MP1964}, and it is commonly 
applied to the {real- or imaginary- time evolution}~\cite{Heller1976,Beck20001,Haegeman2011,Carleo2012,Haegeman2013,Ido_PRB2015,Cevolani2015,Lanata2015,Takai_JPSJ2016,Czarnik2016,Czarnik2016a}. 
From the TDVP, we obtain the following equation
\begin{align}
\sum_{k}\dot{\alpha}_k{\rm Re}\braket{\partial_{\alpha_k}\bar{\psi}|\partial_{\alpha_m}\bar{\psi}}=-{\rm Re}\braket{\bar{\psi}|({H}-\langle{H}\rangle)|\partial_{\alpha_m}\bar{\psi}}.
\end{align}
By discretizing the derivative $\dot{\alpha}_k$ as $\Delta\alpha_k/\Delta \tau$,
we obtain the formula for updating the variational parameters as
\begin{align}
\Delta \alpha_k=-\Delta \tau \sum_{m}S_{km}^{-1}g_{m},
\label{eq:SR_equation_system}
\end{align}
where
\begin{eqnarray}
S_{km}&\equiv&{\rm Re}\braket{\partial_{\alpha_k}\bar{\psi}|\partial_{\alpha_m}\bar{\psi}}\nonumber\\
&=&{\rm Re}\langle {{O}_k^* {O}_m}\rangle-{\rm Re}\langle {{O}_k} \rangle{\rm Re}\langle{{O}_m}\rangle
\end{eqnarray}
and
\begin{eqnarray}
g_{m}&\equiv&{\rm Re}\braket{\bar{\psi}|({H}-\langle{H}\rangle)|\partial_{\alpha_m}\bar{\psi}}\nonumber\\
&=&{\rm Re}\langle {{H} {O}_m}\rangle-\langle {{H}} \rangle{\rm Re}\langle{{O}_m}\rangle.
\end{eqnarray}
$O^*$ means the complex conjugate of $O$.
The operator ${O}_k$ is defined as
\begin{align}
{O}_k=\sum_x\ket{x}\left(\frac{1}{\braket{x|\psi}}\frac{\partial}{\partial \alpha_k}\braket{x|\psi}\right)\bra{x},
\end{align}
where $\ket{x}$ is a real space configuration of electrons.
Here we note that the set of the real 
space configurations $\{\ket{x}\}$ is orthogonal and complete.

To calculate $O_{k}$, it is necessary to
differentiate {the} inner product $\braket{x|\psi}$ with respect 
to 
{a}
variational parameter $\alpha_{k}$.
When  $\alpha_{k}$ is the variational parameter 
of the correlation factor,
it is easy to perform the differentiation.
For example, if $\alpha_{k}$ is the Gutzwiller factor 
at $k$th site
($\alpha_{k}=g_{k}$),
$\braket{x|\psi}$ becomes $e^{g_{k}D_{k}(x)}\braket{x|\psi^{\prime}}$, where
$D_{k}(x)$ is doublon at $k$th site and $\ket{\psi^{\prime}}$ 
is the wavefunction without Gutzwiller factors $g_{k}$.
Thus,  the differentiation becomes
\begin{align}
\frac{\partial\braket{x|\psi}}{\partial g_{k}}
=D_{k}(x)\braket{x|\psi}.
\end{align}

When $\alpha_{k}$ is the variational parameters for the
Pfaffian wave functions,
differentiation of the Pfaffian ${\rm Pf}(X)$ is necessary.
Because the coefficient $F_{AB}$ of the Pfaffian wave function
is a complex number \tohgoe{[$F_{AB}=F_{AB}^{R}+{\rm i}F_{AB}^{I}$, ${\rm Re}(F_{AB})=F_{AB}^{R}$,  ${\rm Im}(F_{AB})=F_{AB}^{I}$]},
it is necessary to consider the differentiations
with respect to both the real part of $F_{AB}$ and the imaginary part of $F_{AB}$.
By using the following formula for the Pfaffian
\begin{align}
\frac{\partial {\rm Pf}[A(x)]}{\partial x}=
\frac{1}{2}{\rm Pf}[A(x)]{\rm Tr}\Big[A^{-1}\frac{\partial A(x)}{\partial x}\tido{\Big]},
\label{Eq:DiffPf}
\end{align}
we obtain 
\begin{align}
\frac{\partial {\rm Pf}(X)}{\partial F_{AB}^{R}}&=-{\rm Pf}(X)(X^{-1})_{AB},\\
\frac{\partial {\rm Pf}(X)}{\partial F_{AB}^{I}}&=-{\rm i}\times{\rm Pf}(X)(X^{-1})_{AB}.
\end{align}
Detailed calculations including the proof of Eq.~(\ref{Eq:DiffPf}) are
shown in \ref{sec:diffPf}.

For the spin-projected wavefunctions, coefficients of $F_{IJ}$ is given by
\begin{align}
F_{IJ}&=f_{ij}\times \tmorita{\kappa(\sigma_i,\sigma_j)}.
\end{align}
In this case, differentiation with respect to $f_{ab}$ are given as
{
\begin{align}
\frac{\partial {\rm Pf}(X)}{\partial f_{ab}^{R}}&=-{\rm Pf}(X)\sum_{\sigma,\sigma^{\prime}}\kappa(\sigma,\sigma^{\prime})(X^{-1})_{a\sigma,b\sigma^{\prime}},\notag \\
\frac{\partial {\rm Pf}(X)}{\partial f_{ab}^{I}}&=-{\rm i}\times{\rm Pf}(X)\sum_{\sigma,\sigma^{\prime}}\kappa(\sigma,\sigma^{\prime})(X^{-1})_{a\sigma,b\sigma^{\prime}}.
\label{eq:diffij}
\end{align}
}
Detailed calculations are given in  \ref{sec:diffPf}.

\subsubsection{Conjugate gradient (CG) method}
\label{sec:CG}
The overlap matrix $S$ is positive definite and 
thus a linear equation system (\ref{eq:SR_equation_system}) can be 
solved by using a Cholesky decomposition $S = LL^\dagger$, where $L$ is a 
lower triangular matrix~\cite{GolubVanLoan, lapack}.
{In the Cholesky decomposition,
it is necessary to store $S$, whose dimension is $O(N_p^2)$. 
Here, $N_{p}$ is 
\imada{the} size of $S$ matrix, i.e., 
\imada{the} number of variational parameters.
For more than a hundred thousand ($10^5$) parameters,
it is difficult to store $S$ matrix in single node and
the memory cost determines the applicable range of mVMC.
}
Neuscamman and co-workers~\cite{Neuscamman_PRB2012} succeeded 
\tido{in optimizing} 
$10^{5}$ parameters 
{with} the conjugate 
gradient (CG) method, which requires only a matrix-vector product to solve a linear equation system.
They derived 
{a way}
to multiply 
\tg{an arbitrary vector $\vec{v}$} 
{by} 
without storing $S$ itself,
which reduces the numerical cost greatly.
We describe their scheme (SR-CG method) in the following.

In the VMC scheme, expectation values $\braket{O_k}$ and $\braket{O_k^*O_m}$ are calculated as mean values over Monte Carlo samples as
\begin{equation}
  \Braket{O_k} = \frac{1}{N_\text{MC}} \sum_\mu \tilde{O}_{k\mu}
\end{equation}
and
\begin{equation}
  \Braket{O_k^* O_m} = \frac{1}{N_\text{MC}} \sum_\mu \tilde{O}_{k\mu}^* \tilde{O}_{m\mu},
\end{equation}
where $\tilde{O}_{k\mu}$ is a value of $O_k$ of the $\mu$th sample $\ket{x_\mu}$, that is,
\begin{equation}
  \tilde{O}_{k\mu} = \frac{\braket{\psi|O_k|x_\mu}}{\braket{\psi|x_\mu}}.
\end{equation}
By using these form, we can multiply $S$ 
\tg{by} an arbitrary real-valued vector $\vec{v}$ as
\begin{equation}
  y_k = \sum_m S_{km} v_m = z^{(1)}_k - z^{(2)}_k,
\end{equation}
where
\begin{align}
  z^{(1)}_k &= \sum_m {\rm Re}\Braket{O_k^* O_m} \yoshimi{v_m} = {\rm Re}\left[\frac{1}{N_\text{MC}} \sum_\mu \tilde{O}_{k\mu}^* \left(\sum_m \tilde{O}^\top_{\mu m} v_m\right)\right],\\
  z^{(2)}_k &= {\rm Re}\Braket{O_k} \sum_m {\rm Re}\Braket{O_m} v_m.
\end{align}
This product requires $2(N_\text{MC}+1)N_p = O(N_pN_\text{MC})$ 
product,
and needs to store $N_p \times N_\text{MC}$ sized matrix $\tilde{O}$ instead of $N_p \times N_p$ sized matrix $S$.
Once a matrix-vector product $\vec{y}=S\vec{v}$ is able to be calculated, 
we can also solve a linear equation system (\ref{eq:SR_equation_system}) by the well-known CG method (Algorithm~\ref{alg:CG}).
Since one iteration includes one matrix-vector product with $O(N_p N_\text{MC})$ products and $5N_p$ products,
and the CG method converges within $N_p$ iterations,
the computational cost of SR-CG in the worst case is $O(N_p^2 N_\text{MC}).$
Therefore, the SR-CG algorithm reduces the numerical cost if $N_\text{MC} < N_p$.
For {details} of the CG method, the readers 
{are referred} 
\tido{to} standard numerical linear algebra textbooks, 
{for example, Ref.~\cite{GolubVanLoan}.}

\begin{algorithm}[tb]
  \caption{CG method for linear equation system $A\vec{x}=\vec{b}$ 
  }
  \begin{algorithmic}[1]
    \Procedure{CG}{$A,\vec{b},\vec{x}_0,\varepsilon^2,k_\text{max}$}
      \State $\epsilon \gets \varepsilon^2\|\vec{b}\|_2^2$ \Comment{termination threshold}
      \State $\vec{x} \gets \vec{x}_0$ \Comment{initial guess}
      \State $\vec{r} \gets \vec{b} - A\vec{x}$ \Comment{residue vector}
      \State $\vec{p} \gets \vec{r}$ \Comment{search direction}
      \State $\rho \gets \|\vec{r}\|_2^2$
      \For{$k \gets 1\dots k_\text{max}$} \Comment{$k_\text{max}$: maximum \# of iterations}
        \State $\vec{w} \gets A\vec{p}$
        \State $\alpha \gets \rho / \left(\vec{p}^\top \vec{w}\right)$
        \State $\vec{x} \gets \vec{x} + \alpha \vec{p}$
        \State $\vec{r} \gets \vec{r} - \alpha \vec{w}$
        \State $\rho' \gets \|\vec{r}\|_2^2$
        \If{$\rho' < \epsilon$}
          \State \textbf{break}
        \EndIf
        \State $\vec{p} \gets \vec{r} + \left(\rho'/\rho\right)\vec{p}$
        \State $\rho \gets \rho'$
      \EndFor
      \State \textbf{return} $\vec{x}$
    \EndProcedure
  \end{algorithmic}
  \label{alg:CG}
\end{algorithm}

\subsection{Power Lanczos method}

In the power-Lanczos method~\cite{Heeb_ZPhys1993}, by multiplying 
\tg{the wave functions by the Hamiltonian}, we systematically improve 
the accuracy of the wave functions.
We explain the basics of the power-Lanczos method in 
\tohgoe{this} subsection.

The wave function with the $n$th-step Lanczos iterations is defined as 
\begin{align}
  \ket{\psi_{\tido{N}}}=(1+\sum_{n=1}^{N}\alpha_{n}H^{n})\ket{\psi},
\end{align}
where $\alpha_{n}$ is a kind of 
variational parameter. 
After the $n$th-step Lanczos iterations, 
$\alpha_{n}$ are determined by minimizing the energy as follows
\begin{align}
  \min_{\vec{\alpha}} E_{\tido{N}}=\tido{\min_{\vec{\alpha}}}\frac{\braket{\psi_{\tido{N}}|H|\psi_{\tido{N}}}}{\braket{\psi_{\tido{N}}|\psi_{\tido{N}}}},
\end{align}
where $\vec{\alpha}=(\alpha_{1},\alpha_{2},\dots,\alpha_{N})$.

In mVMC, the first-step Lanczos iteration is implemented and
the energy is calculated as
\begin{align}
E_{1}(\alpha_{1}) =\frac{\braket{\psi_{1}|H|\psi_{1}}}{\braket{\psi_{1}|\psi_{1}}}
=\frac{h_1 + \alpha_{1}(h_{2(20)} + h_{2(11)}) + \alpha_{1}^2 h_{3(12)}}{1 + 2\alpha_{1} h_1 + \alpha_{1}^2 h_{2(11)}},
\end{align}
where we define $h_1$, $h_{2(11)},~h_{2(20)},$ and $h_{3(12)}$ as
\begin{align}
&h_1 =\sum_{x} \rho(x) F^{\dag}(x,  {H}),\\
&h_{2(11)}=\sum_{x} \rho(x) F^{\dag}(x,  {H}) F(x, {H}),\\
&h_{2(20)}=\sum_{x} \rho(x) F^{\dag}(x,  {H}^2),\\
&h_{3(12)}=\sum_{x} \rho(x) F^{\dag}(x,  {H})F(x,  {H}^2),\\
& \rho(x)=\frac{|\braket{\psi|x}|^2}{\braket{\psi|\psi}},~F(x,{A})=\frac{\braket{x|A|\psi}}{\braket{x|\psi}}.
\end{align}
From the condition 
\begin{align}
\frac{\partial E_{1}(\alpha_{1})}{\partial \alpha_{1}}=0, 
\end{align}
i.e., by solving the quadratic equations, we can determine the optimized {value of the parameter} $\alpha_{1}$. 
By using the optimized $\alpha_{1}$, we can calculate 
other expected values in a similar way.

\begin{figure}[tb!]
 \begin{center}
  \includegraphics[width=6.5cm]{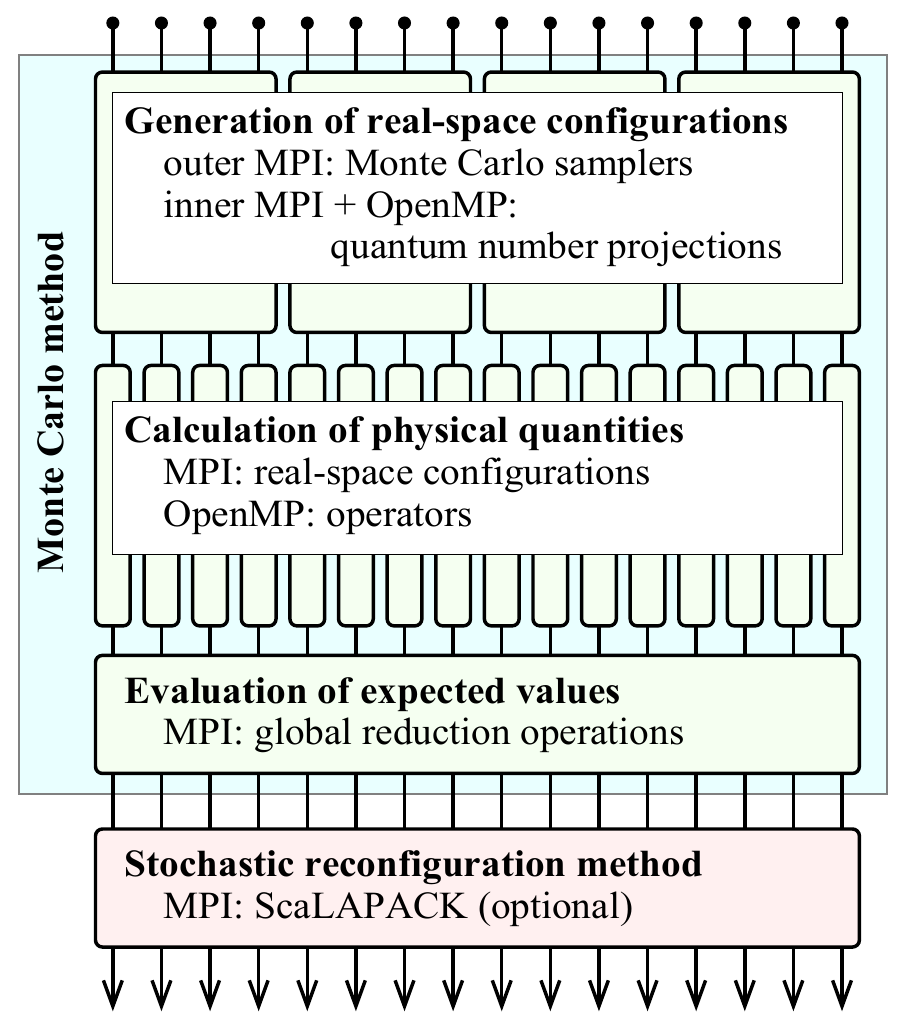}
  \caption{Parallelization of mVMC calculations.  The vertical lines
  indicate MPI processes.  In this figure, four Monte Carlo samplers
  generate the real-space configurations and each sampler is
  parallelized with four MPI processes.}
  \label{fig:parallel}
 \end{center}
\end{figure}

\subsection{Parallelization}
\label{sec:Para}

mVMC supports MPI/OpenMP hybrid parallelization.  We adopt different
parallelization approaches in the Monte Carlo method and the
optimization method as shown {in} Fig.\ref{fig:parallel}.

In the Monte Carlo method, $N_\text{sampler}$ independent Monte Carlo
samplers are created.  The number of MPI processes per sampler is unity
by default but can be specified from the input file.  First each sampler
generates $N_\text{MC}/N_\text{sampler}$ real-space configurations
$\{|x\rangle\}$.  In each MC sampler, the iterations of loops for
summations in the quantum-number projections
(\ref{Eq:momP},\ref{Eq:pointP},\ref{Eq:spnP}) are parallelized using
both MPI and OpenMP.  Then the real-space configurations are distributed
among MPI processes and physical quantities
$\langle\psi|A|x\rangle/\langle\psi|x\rangle$ are calculated
independently.  The operators $\{A\}$ are distributed among OpenMP
threads here.  Finally, summations over the Monte Carlo samples
(\ref{Eq:MC}) are executed by global reduction operations with
collective communication.

In the SR method, if the CG method is not used, the linear equation
(\ref{eq:SR_equation_system}) is solved by using ScaLAPACK routines. The
elements of the overlap matrix $S$ are distributed in block-cyclic
fashion.  In the CG method, multiplication between the matrix $S$ and a
vector is parallelized in the same manner as {in} the Monte Carlo method.

\section{Benchmark results}
In this section, 
we show benchmark results of mVMC calculations
for several standard models in the condensed matter physics
such as the Hubbard model,
the Heisenberg model and the Kondo-lattice model. 
{For these models, we compare the results by
mVMC 
\imada{with} the results \imada{either} by the exact diagonalization 
\imada{or} by the unbiased quantum Monte Carlo calculations.
From these comparisons,
we show that mVMC can generate highly accurate
wave functions for the ground states and
the low-energy excited states in
these standard models.
}

\subsection{Hubbard  model and Heisenberg model on the square lattice}
We show examples of the mVMC calculations for
the Hubbard and Heisenberg {models} on the square lattice.
We compare the mVMC calculation 
with the exact diagonalization and the available 
unbiased quantum Monte Carlo calculations in the literature.

\subsubsection{Comparison with exact diagonalization}
\begin{figure}[tb!]
  \begin{center}
    \includegraphics[width=7.5cm]{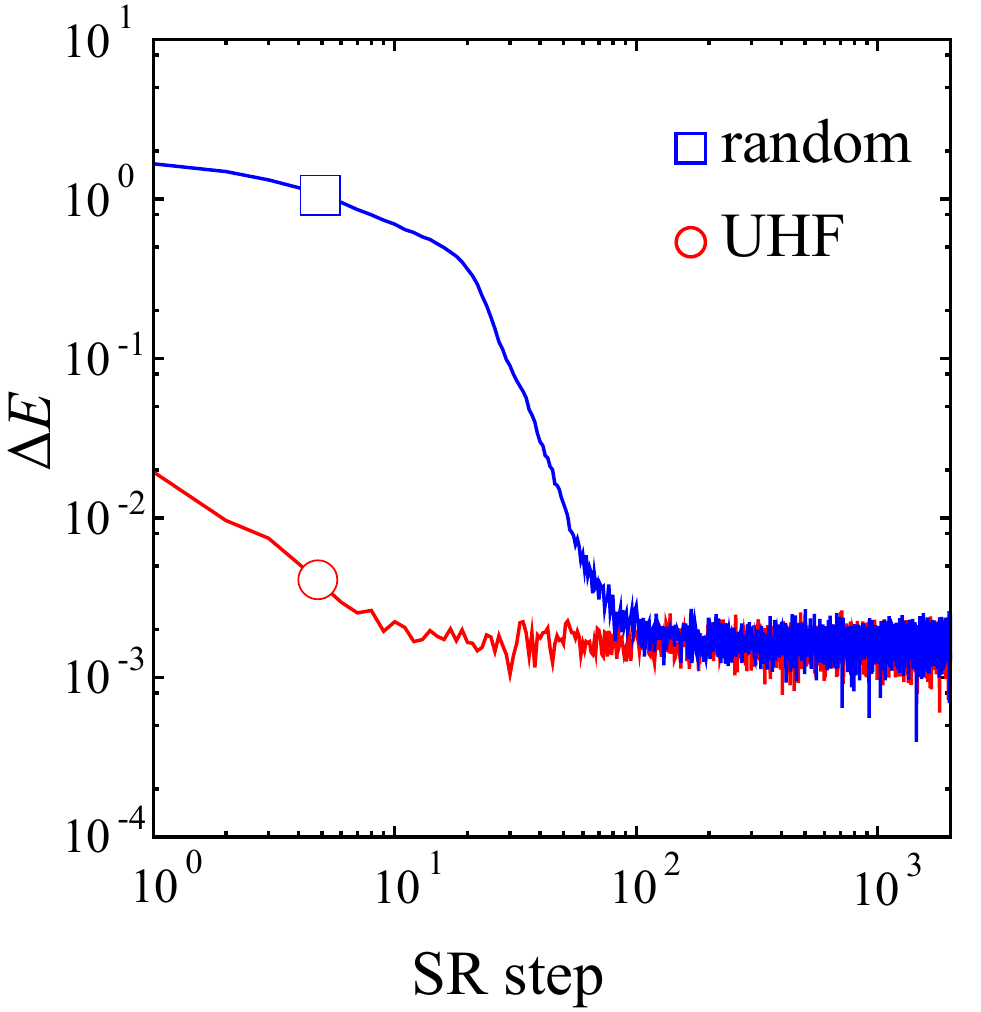}
    \caption{Optimization processes for $4\times4$ Hubbard model for $U=4$ and $t=1$ at half filling. 
By taking the 
UHF solutions as an initial state, one can reach the ground state 
faster than the random initial states.}
    \label{fig:SR}
  \end{center}
\end{figure}

By using Standard mode in mVMC, it is easy to
start the calculation for the Hubbard model.
An example of the input file for 
$4\times4$-site Hubbard model at half filling is shown as follows:
\begin{verbatim}
model         = "FermionHubbard"
lattice       = "square"
W             = 4
L             = 4
Wsub          = 2
Lsub          = 2
t             = 1.0
U             = 4.0
nelec         = 16
2Sz           = 0
NMPTrans      = 4
\end{verbatim}

In Fig.~\ref{fig:SR}, 
we show typical optimization processes for 
the Hubbard model.
We take two initial states in mVMC calculations;
one is \yoshimi{a} random initial state, i.e., 
each variational parameter of the 
{pair-product} part is
given by 
pseudo random numbers.
Another 
initial state is the {unrestricted Hartree-Fock (UHF)} 
solutions.
Following the relation between the Slater determinant 
and the Pfaffian wave functions [see Eq.(\ref{eq:fijSL})],
we generate the variational parameters of the 
Pfaffian wave function from the Hartree-Fock solutions.
The \yoshimi{program} for the 
{UHF} calculations
is included in mVMC package.
As shown in Fig.~\ref{fig:SR}, 
by taking the
Hartree-Fock solution as an initial state,
optimization becomes faster 
{than} \tohgoe{with} 
{a} random initial 
\tohgoe{state}.
This result shows that 
preparing {a} proper initial state 
is important for efficient optimization.

We examine the accuracy of mVMC calculation by 
comparing 
\yoshimi{to} the exact diagonalization \yoshimi{result}. 
In mVMC calculations, we can improve the accuracy of 
the wave function by extending sublattice structures in
the variational wave functions because 
larger sublattice structures can take into account the long-range fluctuations.
In 
\tido{T}able~\ref{table:L4Hub}, 
we compare several physical 
\tido{quantities} such as 
the 
energy per site $E/N_{\rm s}$,
the doublon density $D$,
the 
nearest-neighbor spin correlations $S_{\rm nn}$,
and the peak value of the spin structure factor $\tilde{S}(\vec{k})$.
Definitions of the physical 
\tido{quantities} are given 
as follows:
{
\begin{align}
D          &=\frac{1}{N_{\rm s}}\sum_{i}\braket{ n_{i\uparrow}n_{i\downarrow}}, \\
S_{\rm nn} &=\frac{1}{4N_{\rm s}}\sum_{i,\mu}\braket{\vec{S}_{\vec{r}_{i}}\cdot\vec{S}_{\vec{r}_{i}+\vec{e}_{\mu}}}, \\
\tilde{S}(\vec{k})&=\frac{S({\vec{k})}}{3N_{\rm s}},
\end{align}
}
where $\vec{e}_{\mu}$ denotes the 
nearest-neighbor translational vector, 
{and $S(\vec{k})$ is defined in Eq.~(\ref{eq:Sk}).}
One can see that the \imada{accuracy} 
of the 
energy 
\imada{is}
improved by taking the large sublattice structures. In addition to the energy, 
it is shown that accuracies of 
other physical 
\tido{quantities} are also improved.

{We also perform the first-step power Lanczos calculation,
which can systematically improves the accuracy of the
wave functions.
As a result, we show that the first-step power Lanczos 
{iteration improves}
the accuracy of the energy 
{although other physical properties are not sensitive to the
Lanczos iteration at 
{least} for the system sizes studied here.}
The relative error 
\begin{equation}
\eta=|E_{\rm ED}-E_{\rm mVMC}|/|E_{\rm ED}|
\end{equation}
becomes about $\tido{10^{-4}}$(\tido{$10^{-2}$}\%) 
for the best mVMC calculation [mVMC($4\times4$)+Lanczos].
{This result shows that mVMC can generate highly accurate wavefunctions 
in the Hubbard model.}
}

\begin{table*}[tb!]
\caption{Comparisons with exact diagonalization \yoshimi{(ED)} 
for $4\times4$ Hubbard model with $U=4$ and $t=1$ at half filling.
\yoshimi{ED} is done by using $\mathcal{H}\Phi$~\cite{hphi,hphi_ma}.
mVMC($2\times2$\tido{/$4\times4$}) means $f_{ij}$ has \tido{the} $2\times2$\tido{/$4\times 4$} sublattice structure.
$N_p$ is \imada{the} 
size of $S$ matrix, \imada{which is the number of variational parameters} 
and $\vec{k}_{\rm peak}=(\pi,\pi)$.
\imada{Error bars  are denoted by the parentheses in the last digit.}
Lanczos means that the first-step Lanczos calculations on top of the
mVMC calculations.
{In the Lanczos calculations, to reduce the numerical cost, 
we calculate the diagonal spin correlations such as
$S^{z}_{\rm nn}=3/4N_{\rm s}\sum_{i,\mu} \langle S^{z}_{\vec{r}_{i}}\cdot S^{z}_{\vec{r}_{i}+\vec{e}_{\mu}}\rangle$
and
$\tilde{S}^{z}(\vec{k})=S^{z}(\vec{k})/N_{\rm s}$, which are
equivalent to $S_{\rm nn}$ and $\tilde{S}(\vec{k})$ 
when the spin-rotational symmetry is preserved.}
}
\begin{center}
\begin{tabular}{llllll}
\hline 
                         & $E/N_{\rm s}$   & $D$         & $S_{\rm nn}$ & $\tilde{S}(\vec{k}_{\rm peak})$ & {$N_{p}$}  \\ \hline
ED                       & -0.85136        &  0.11512    & -0.2063      & 0.05699                       & -            \\
mVMC($2\times2$)         & -0.84985(3)     &  0.1155(1)  & -0.2057(2)   & 0.05762(4)                    & 74           \\
mVMC($2\times2$)+Lanczos & -0.85100(2)     &  0.1156(1)  & -0.2054(8)   & 0.05736(2)                    & 74           \\
mVMC($4\times4$)         & -0.85070(2)     &  0.1151(1)  & -0.2065(1)   & 0.05737(2)                    & 266          \\
mVMC($4\times4$)+Lanczos & -0.85122(1)     &  0.1151(1)  & -0.2072(4)   & 0.0576(1)                     & 266          \\
\hline
\end{tabular}
\end{center}
\label{table:L4Hub}
\end{table*}

Next, in Table  \ref{table:L4Hei}, we show the results for the
Heisenberg model on the square lattice,
which is a strong coupling limit of the 
Hubbard model at half filling.
An example of the input file for 
the $4\times4$-site Heisenberg model is shown as follows:
\begin{verbatim}
model         = "Spin"
lattice       = "square"
W             = 4
L             = 4
Wsub          = 2
Lsub          = 2
J             = 1.0
2Sz           = 0
NMPTrans      = 4
\end{verbatim}
Because the exact diagonalization can be performed up to $6\times 6$ {square lattice}
within the realistic numerical cost,
we compare mVMC calculations 
\yoshimi{to} exact diagonalization \yoshimi{results} for
\tido{the} $4\times 4$ and $6\times 6$ Heisenberg model. In addition to the energy,
we calculate the nearest-neighbor spin correlations $S_{\rm nn}$,
the next-nearest-neighbor spin correlations $S_{\rm nnn}$, and
the spin structure factors $\tilde{S}(\vec{k}_{\rm peak})$.
Definitions of the physical 
\tido{quantities} are the same as those 
of the Hubbard model.
As {it} is shown in the Hubbard model,
by taking 
{a}
large sublattice structure
and performing the power Lanczos method,
the accuracies of the wave functions are improved.
For the best mVMC calculations,
the relative error $\eta$ becomes
\tido{$10^{-8}$}(\tido{$10^{-6}$}\%) for $L=4$, and
\tido{$10^{-5}$}(\tido{$10^{-3}$}\%) for $L=6$.
For $L=4$, because the Hilbert space is small (about 16000), 
mVMC with $4\times4$ sublattice structure gives nearly
{the} exact ground-state energy.
In this situation,
the power-Lanczos calculations become unstable 
because the variance becomes almost zero.
Thus, we do not show the results of the power-Lanczos method
{for mVMC($4\times4$)}
in \tido{T}able \ref{table:L4Hei}.

\begin{table*}[tb!]
\caption{Comparisons with exact diagonalization for  $4\times4$ and $6\times6$ Heisenberg model 
with $J=1$. 
We note $\vec{k}_{\rm peak}=(\pi,\pi)$.
The relative errors $\eta$ become
$10^{-6}$\% for $L=4$ and $10^{-3}$\% for $L=6$, respectively. 
{The definitions of the spin correlations in the Lanczos method 
\imada{and $N_{p}$} are same as those 
of Table \ref{table:L4Hub}.}}
\begin{center}
\begin{tabular}{llllll}
\hline 
($N_{\rm s}=4\times4$)    & $E/N_{\rm s}$   & $S_{\rm nn}$   & $S_{\rm nnn}$ & $\tilde{S}(\vec{k}_{\rm peak})$  & {$N_{p}$} \\ \hline
ED                        & -0.70178020     & -0.35089010    & 0.21376       & 0.09217                          & -                \\
mVMC($2\times2$)          & -0.701765(2)    & -0.350883(1)   & 0.2136(1)     & 0.09216(3)                       & 64               \\ 
mVMC($2\times2$)+Lanczos  & -0.701780(1)    & -0.3517(5)     & 0.214(1)      & 0.0924(2)                        & 64               \\ 
mVMC($4\times4$)          & -0.70178015(8)  & -0.35089007(4) & 0.2139(4)     & 0.0922(1)                        & 256              \\
\hline 
($N_{\rm s}=6\times6$)    & $E/N_{\rm s}$   & $S_{\rm nn}$  & $S_{\rm nnn}$ & $\tilde{S}(\vec{k}_{\rm peak})$  & {$N_{p}$}  \\ \hline
ED                        & -0.678872       & -0.33943607   & 0.207402499   & 0.069945                         & -                 \\
mVMC($2\times2$)          & -0.67846(1)     & -0.33923(1)   & 0.20742(3)    & 0.07021(2)                       & 144               \\ 
mVMC($2\times2$)+Lanczos  & -0.678840(4)    & -0.339(1)     & 0.207(1)      & 0.0698(3)                        & 144               \\ 
mVMC($6\times6$)          & -0.678865(1)    & -0.3394326(4) & 0.20735(4)    & 0.06993(3)                       & 1296              \\
mVMC($6\times6$)+Lanczos  & -0.678871(1)    & -0.3391(5)    & 0.2071(6)     & 0.0699(2)                        & 1296              \\
\hline 
\end{tabular}
\end{center}
\label{table:L4Hei}
\end{table*}

\begin{figure}[tb!]
  \begin{center}
    \includegraphics[width=7.5cm]{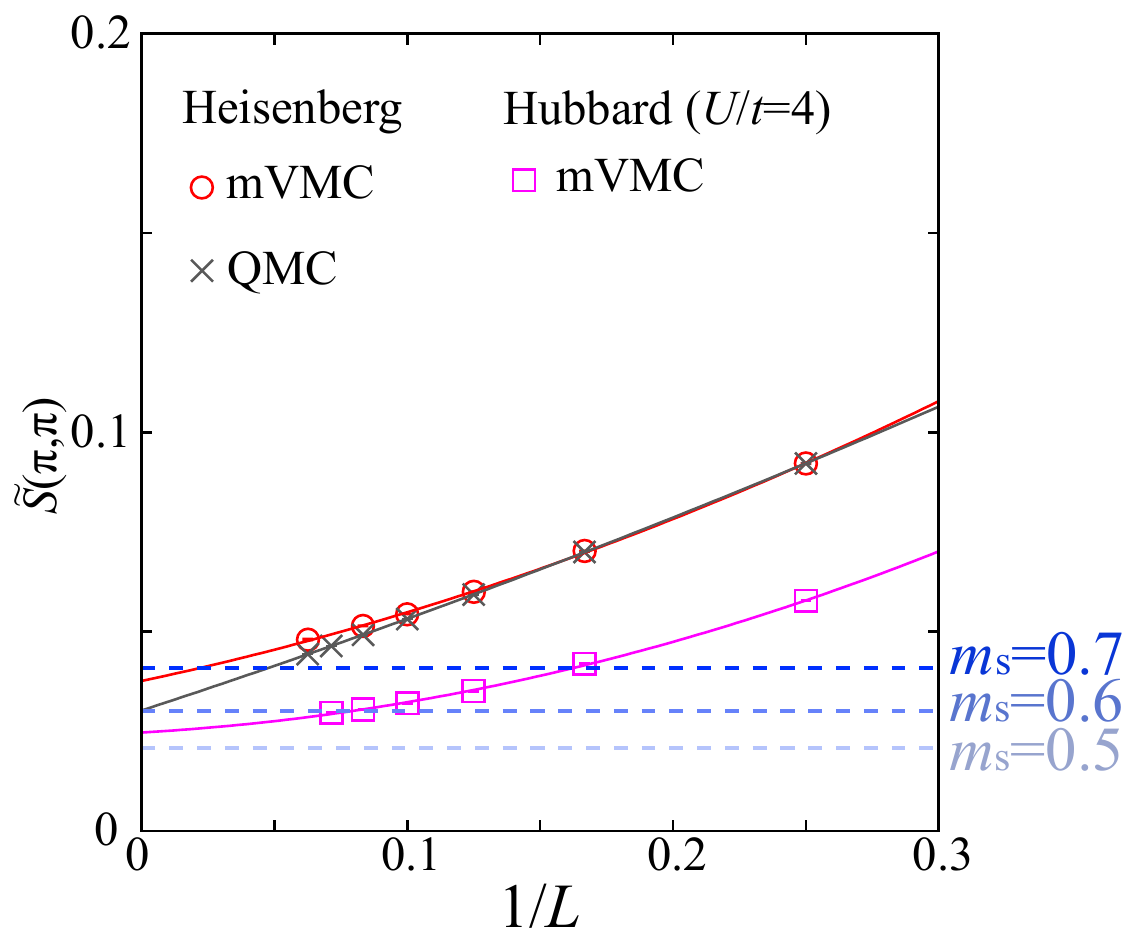}
    \caption{ Size-extrapolation of the spin structure factors
for the Hubbard model with $U/t=4$ on the square lattice 
and the Heisenberg model on the square lattice.
{We perform size extrapolation by fitting data with the second-order polynomials.}
    }
    \label{fig:Sq}
  \end{center}
\end{figure}

To examine the accuracy of mVMC beyond the system size that can be 
treated by the exact diagonalization,
we calculate \tido{the} $L\times L$ Hubbard and 
Heisenberg 
\yoshimi{models}.
For large system sizes, the numerical cost becomes large,
we only perform the $2\times 2$ sublattice calculations and
do not perform the power-Lanczos calculations.
In 
\tohgoe{Fig.} \ref{fig:Sq}, we show \yoshimi{the} size dependence of the 
$\tilde{S}(\pi,\pi)/N_{\rm s}$ for the Hubbard model with $U/t=4$
and the Heisenberg model.
It has been shown that the \tohgoe{ground states} of the Hubbard model on the 
square lattice and the Heisenberg model on the square lattice
have long-range antiferromagnetic order.
To detect the  antiferromagnetic order, we perform the 
size-extrapolation of the spin 
\tohgoe{structure} factors and obtain the
spontaneous magnetization $m_{s}$, which is given by
\begin{align}
m_{s} = 2\Big[\lim_{L\rightarrow\infty} 3\tilde{S}(\pi,\pi)\Big]^{1/2}.
\end{align}
We note that the full saturated magnetic moment is given by $m_{s}=1$ in this definition.

For the Heisenberg model, 
we plot results 
{from} the 
unbiased quantum Monte Carlo method in 
\tido{R}ef.~\cite{Sandvik_PRB1997} and
compare with mVMC calculations.
For $L\leq 10$, the agreement is nearly perfect, i.e.,
mVMC can well reproduce the results by the QMC calculations
for larger system sizes beyond the exact diagonalization.
We find that
the deviation from 
QMC results becomes larger
for $L\geq 12$.
We estimate $m_{s}=0.67$ from mVMC calculations, which is 
higher than the estimate by the QMC calculations
($m_{s}=0.607$).
This deviation originates from the 
\tohgoe{lower}
accuracy for the larger system sizes.
By taking large-sublattice structure and
performing power-Lanczos calculation, one can resolve this slight
deviations.
However, because such extended calculations are expensive 
and beyond the scope of this paper,
we do not perform such calculations. 
For the Hubbard model with $U/t=4$, 
we obtain $m_{\rm s}=0.54$
which is again slightly larger than 
those {
{from} the} QMC calculations 
($m_{\rm s}\sim0.48$~\cite{Varney_PRB2009,Tahara_JPSJ2008}).

\subsection{Kondo lattice model on the one-dimensional chain }
We apply mVMC to the Kondo lattice model in which 
the local spin-$1/2$ spins
couple with the 
\tohgoe{conduction} electrons through the
Kondo coupling $J$~\cite{Tsunetsugu_RMP1997}.
In \tohgoe{mVMC}, we describe the local spin by completely excluding the 
double occupancy.
To examine the accuracy of mVMC for the Kondo lattice model,
we perform mVMC calculations for one-dimensional Kondo lattice model.
An example of the input file for 
the $8$-site Kondo-lattice model is shown as follows:
\begin{verbatim}
model         = "Kondo"
lattice       = "chain"
L             = 8
Lsub          = 2
t             = 1.0
J             = 1.0
nelec         = 8
2Sz           = 0
NMPTrans      = 2
\end{verbatim}

\begin{table*}[tb!]
\caption{
Comparisons with exact diagonalization for one-dimensional Kondo-lattice model
with $J=1$, $t=1$, and $L=8$. 
\imada{Notations are the same as Table 3.}
Upper (Lower) panel shows the results for spin singlet (triplet) sector.
In the triplet sector ($S=1$), we take total momentum as $K=\pi$, which gives the lowest 
energy in $S=1$, while we take 
total momentum as $K=0$ for $S=1$.
{The definitions of the spin 
correlations in the Lanczos method 
\imada{are} {the} same as those of Table \ref{table:L4Hub} for $S=0$.
For $S=1$, because spin-rotational symmetry is not preserved 
  and $S^{z}$ correlations 
 {are} not equivalent to that of $S^{x}$ and $S^{y}$ correlations.
{We} do not show the results of the spin correlations in the Lanczos method for $S=1$.}
}
\begin{center}
\begin{tabular}{llcccc}
\hline 
($L=8$, $S=0$)              & $E/N_{\rm s}$   & $S_{\rm onsite}$ & $S_{\rm nn}^{\rm loc}$  & $S(\pi)$   & {$N_{p}$} \\ \hline
ED                   & -1.394104       & -0.3151          & -0.3386                 & 0.05685           & -             \\
mVMC($2$)            & -1.39350(1)     & -0.3144(1)       & -0.3363(1)              & 0.05752(3)        & 69             \\ 
mVMC($2$)+Lanczos    & -1.39401(2)     & -0.3152(2)       & -0.336(1)               & 0.05716(4)        & 69             \\ 
mVMC($8$)            & -1.39398(1)     & -0.3151(2)       & -0.3384(2)              & 0.05693(4)        & 261             \\ 
mVMC($8$)+Lanczos    & -1.394097(2)    & -0.3150(2)       & -0.3377(3)              & 0.0568(1)         & 261             \\ 
\hline \hline
($L=8$, $S=1$)              & $E/N_{\rm s}$  & $S_{\rm onsite}$ & $S_{\rm nn}^{\rm loc}$  & $S(\pi)$    & {$N_{p}$}\\ \hline
ED                   & -1.382061      & -0.2748          & -0.2240                 &  0.05747           & - \\
mVMC($2$)            & -1.38126(3)    & -0.2738(2)       & -0.2246(1)              &  0.05822(1)        & 69 \\ 
mVMC($2$)+Lanczos    & -1.38187(1)    & -                & -                       &  -                 & 69 \\ 
mVMC($8$)            & -1.38171(3)    & -0.2750(4)       & -0.2249(7)              &  0.0577(1)         & 261\\ 
mVMC($8$)+Lanczos    & -1.382011(2)   & -                & -                       &  -                 & 261 \\ 
\hline 
\end{tabular}
\end{center}
\label{table:Kondo}
\end{table*}

In 
\tido{T}able \ref{table:Kondo}, we compare mVMC calculations
{to} the exact diagonalization {results}.
In \tido{the} mVMC calculations, we examine the \yoshimi{size dependence of} sublattice structure and
\yoshimi{the improvement by} the power-Lanczos method.
{Equally to the previous cases,}
\tido{the} mVMC calculations {for the Kondo lattice model} well reproduce
the result\tido{s} of the exact diagonalization and 
the relative error becomes \tido{$10^{-4}$}\%.
Other physical 
\tido{quantities} such as 
the local spin correlations ($S_{\rm onsite}$) and 
next-neighbor spin correlations for 
local spins ($S^{\rm loc}_{nn}$),
and the spin structures are 
well reproduced as shown in \tido{T}able \ref{table:Kondo}.
Definitions of the physical \tido{quantities} are given as
\begin{align}
S_{\rm onsite}&=\frac{1}{N_{s}}\sum_{i}\langle\vec{S}_{i}\cdot\vec{s}_{i}^{c}\rangle, \\
S_{\rm nn}^{\rm loc}&=\frac{1}{4N_{s}}\sum_{i,\mu}\langle\vec{S}_{\vec{r}_{i}+\vec{e}_{u}}\cdot\vec{S}_{\vec{r}_{i}}\rangle,
\end{align}
where $\vec{S}_{i}$ ($\vec{s}_{i}$) denotes
the spin operators for local (conduction electrons) spins.

In addition to the \tohgoe{ground state},
by selecting the different 
quantum number, it is possible
to obtain the low-energy state   
by using mVMC.
In the lower panel in 
\tido{T}able \ref{table:Kondo},
we show the results of the spin triplet sector.
We note that the total momentum of the 
lowest-energy state in the
spin triplet sector is given by $K=\pi$.
As show{n} in 
\tido{T}able \ref{table:Kondo},
mVMC well reproduces the low-energy-excited state. 

\section{Summary}
In summary,
we {have exposited} the basic usage of mVMC in Sec.~2 
such as {the way} 
to download {and build} mVMC and how to begin 
calculations by mVMC.
In Standard mode, by preparing one input file whose
length is typically less than ten lines,
one can perform mVMC calculations for
standard models in the condensed matter physics
such as the Hubbard model, the Heisenberg model,
and the Kondo-lattice model. 
By changing {keywords} for lattice,
one can treat several lattice structures
such as the one-dimensional chains, 
the square lattice, the triangular lattice, and so on.
We 
{have also presented}
the visualization tools included 
in mVMC package.

In Sec.3, we 
{have explained}
the basic algorithms used 
in mVMC calculations such as {the} sampling method and
the update techniques in the sampling.
We 
have also detailed the 
wavefunctions implemented  in mVMC and 
the quantum number projections. 
In Sec.3.4,
we 
{have expounded on the }
optimization method used in mVMC, i.e.,
the basics of the
SR method.
\tido{In order to solve the large linear equations in the SR method, 
we employ the efficient algorithm proposed 
by Neuscamman {\it et al.} (SR-CG method)~\cite{Neuscamman_PRB2012}.
The SR-CG {method} is detailed in Sec.3.4.2.}
{Although the SR-CG method is an efficient method 
for optimizing a large number of parameters, 
\imada{more efficient method for smaller number of variational parameters ($<$2000) 
were proposed~\cite{Toulouse_JCP2008}. 
It is an intriguing future issue to examine it for the small size calculations.}}

In Sec. 4, we {have shown} several {benchmark results} of mVMC calculations.
For the Hubbard \tido{model}, Heisenberg model, and the Kondo-lattice \tido{model},
we {have compared} mVMC calculations 
\yoshimi{to} the exact diagonalization \yoshimi{results}
and available unbiased calculations in the literature.
For all the models, we 
{have shown} that mVMC
well reproduces the results by the exact diagonalization 
and the unbiased calculation\tido{s} by the QMC.
We note that mVMC well reproduce\tido{s} not only energy but also
other physical 
\tido{quantities} such as the spin correlations.
For the Kondo lattice model, we {have presented} the low-energy excited
state by choosing the spin triplet sector through
the quantum number projections and
{have shown} that mVMC also reproduces the low-energy excited state\tido{s}
with high accuracy.
{All results show that  mVMC generates highly accurate
wavefunctions for}
\tido{the quantum many-body systems}.
 
Recent studies show that the accuracy of 
the 
VMC method is much more improved by 
introducing the backflow correlations for Pfaffian wavefunctions~\cite{Ido_PRB2015}
and combining with the tensor network method~\cite{Zhao_RPB2017}.  
Furthermore, in addition to the ground-state calculations, 
recent studies show that the 
VMC method can be applicable to
the real-time evolution 
and the finite-temperature calculations 
in the strongly correlated electron systems\tg{\cite{Ido_PRB2015, Takai_JPSJ2016}}.
Implementation of such extensions in mVMC is 
a promising way to make mVMC more useful software and
will be reported in the near future.

\section{Acknowledgements}
We would like to express our sincere gratitude to Daisuke Tahara 
for providing us his code of variational Monte Carlo method.
A part of mVMC is based on his original code.
We also acknowledge Hiroshi Shinaoka, Youhei Yamaji, 
Moyuru Kurita, Ryui Kaneko, and Hui-Hai Zhao for their cooperations on developing mVMC.
We would also like to thank the support from 
``{\it Project for advancement of software usability in materials science}" 
by Institute for Solid State Physics, University of Tokyo, 
for development of mVMC ver.1.0.
This work was also supported by Grant-in-Aid for 
Scientific Research (16H06345 and 16K17746)
and Building of Consortia for the Development of Human Resources
in Science and Technology from the MEXT of Japan.
We also thank numerical resources from the Supercomputer 
Center of Institute for Solid State Physics, University of Tokyo.
KI was financially supported by Grant-in-Aid for JSPS Fellows (No. 17J07021) and 
Japan Society for the Promotion of Science through Program for Leading Graduate Schools (MERIT).
{
We thank
the computational resources of the K computer provided
by the RIKEN Advanced Institute for Computational
Science through the HPCI System Research project, as
well as the project "Social and scientific priority issue 
(Creation of new functional devices and high-performance materials
to support next-generation industries; CDMSI)" to
be tackled by using post-K computer, under the project
number hp160201, and hp170263 supported by Ministry of Education, Culture,
Sports, Science and Technology, Japan (MEXT) .
}

\appendix

\section{Anti-periodic boundary conditions}
\label{sec:anti}
In mVMC, we can specify the phase for the hopping through the cell boundary
  with \verb|phase0| {($\vec{a}_0$ direction)} and \verb|phase1| {($\vec{a}_1$ direction)}.
  For example, a hopping from the $i$th site to the $j$th site 
  through the cell boundary with the positive $\vec{a}_0$ direction becomes 
  \begin{align}
    &\exp({\rm i} \times {\rm phase0}\times\pi/180) 
    \times t {c}_{j \sigma}^\dagger {c}_{i \sigma} \nonumber \\
    &+ \exp(-{\rm i} \times {\rm phase0}\times\pi/180) 
    \times t^* {c}_{i \sigma}^\dagger {c}_{j \sigma}.
  \end{align}
  By using \verb|phase0| and \verb|phase1| we can treat the anti-periodic boundary conditions.
  For example, by taking \verb|phase0|=$\pi$ in the one-dimensional chain,
  we can treat the anti-periodic boundary conditions.

When the anti-periodic boundary condition is used,
the sign of the creation/annihilation operator 
changes when the site index is translated across
the boundary of the simulation cell.
This causes the sign change of the pair wavefunction $f_{ij}$ in
the momentum projection or the sub-lattice separation.
For example, we demonstrate the case of one dimensional four-sites system with the
anti-periodic boundary condition.
If we translate a term $f_{0 3} c^\dagger_{0\uparrow} c_{3 \downarrow}$ with +2,
this term becomes 
\begin{align}
T_{+2} f_{0 3} c_{0 \uparrow}^\dagger c_{3 \downarrow}^\dagger
= f_{0 3} c_{2 \uparrow}^\dagger c_{5 \downarrow}^\dagger
= - f_{0 3} c_{2 \uparrow}^\dagger c_{1 \downarrow}^\dagger,
\end{align}
where $T_{+2}$ is the translation operator, and
we use the anti-periodic boundary condition at the last equation.
If we divide this system into two sub-lattices,
the wavefunction should be invariable to the translation $T_{+2}$.
Therefore, the variational parameter must satisfy the condition $f_{0 3} = - f_{2 1}$.

In addition, we also should consider the sign-change in the momentum projection. 
The translation $T_{\vec{R}}$ which associate to the vector $\vec{R}$ is applied
to the two-body wavefunction as follows:
\begin{align}
T_{\vec{R}} |\varphi_{\rm pair}\rangle
&= T_{\vec{R}} \left[
\sum_{i,j} f_{ij} c_{i\uparrow}^\dagger c_{j\downarrow}^\dagger
\right]^{N_{\tido{\rm e}}/2} | 0 \rangle
\nonumber \\
&=
\left[
  \sum_{i,j} f_{ij} c_{T_{\vec{R}}(i)\uparrow}^\dagger c_{T_{\vec{R}}(j)\downarrow}^\dagger
\right]^{N_{\tido{\rm e}}/2} | 0 \rangle.
\label{eq:app_antiperiod}
\end{align}
When the site translated from site $i$ with the vector $\vec{R}$ [$T_{\vec{R}}(i)$] crosses the boundary,
it is out of the range from 1 to $N_s$.
We define the site ${\tilde T}_{\vec{R}}(i)$ as a site folded into the original simulation cell
from the site $T_{\vec{R}}(i)$ by using the periodicity and
$s_{\vec{R}}(i) (=\pm 1)$ as a sign caused during the translation from
$T_{\vec{R}}(i)$ to ${\tilde T}_{\vec{R}}(i)$ across anti-periodic boundary, 
i.e.,
$T_{\vec{R}} c_{i} = s_{\vec{R}}(i) c_{{\tilde T}_{\vec{R}} (i)}$.
Then Eq. (\ref{eq:app_antiperiod}) becomes
\begin{align}
T_{\vec{R}} |\varphi_{\rm pair}\rangle
&= \left[\sum_{i,j}
f_{ij} s_{\vec{R}}(i) s_{\vec{R}}(j) c_{{\tilde T}_{\vec{R}}(i)\uparrow}^\dagger
c_{{\tilde T}_{\vec{R}}(j)\downarrow}^\dagger
\right]^{N_{\tido{\rm e}}/2} | 0 \rangle
\nonumber \\
&= \left[\sum_{k,l}
f_{{\tilde T}_{- \vec{R}}(k),{\tilde T}_{-\vec{R}}(l)} s_{-\vec{R}}(k) s_{-\vec{R}}(l)
c_{k\uparrow}^\dagger c_{l\downarrow}^\dagger
\right]^{N_{\tido{\rm e}}/2} | 0 \rangle,
\end{align}
where $k=T_{\vec{R}}(i)$, $l=T_{\vec{R}}(j)$.

In mVMC, we can specify the anti-symmetric condition of 
\imada{two-body} wavefunctions.
(part $f_{i j} = - f_{k l}$) as well as
${\tilde T}_{\vec{R}} (i)$ and $s_{\vec{R}}(i)$ for the quantum number projection.
If we use anti-periodic boundary condition in Standard mode,
they are automatically 
{adjusted without any additional work.}

\section{Relation between {a} Pfaffian 
\tido{wave function} and single Slater determinant}
\label{sec:PfAndSlater}
In this section, we show \tido{the} relation between 
Pfaffian wavefunctions and Slater determinants, which are defined as
\begin{align}
  |\phi_{\rm Pf}\rangle&=\Big(\sum_{I,J=0}^{2N_{\tido{\rm s}}-1}F_{IJ}c_{I}^{\dagger}c_{J}^{\dagger}\Big)^{N_{\rm e}/2}|0\rangle, \\
  |\phi_{\rm SL}\rangle&=\Big(\prod_{n=1}^{N_{\tido{\rm e}}}\psi_{n}^{\dagger}\Big)|0\rangle,~~\psi_{n}^{\dagger}=\sum_{I=1}^{2N_{\tido{\rm s}}}\Phi_{In}c^{\dagger}_{I},
\end{align}
where $I,J$ denote the site index including the spin degrees of freedom.
We assume that $\Phi$ is the normalized orthogonal basis, i.e\tido{.}, 
\begin{equation}
  \sum_{I=0}^{2N_{\tido{\rm s}}-1}\Phi_{In}^{*}\Phi_{Im}=\delta_{nm},
\end{equation}
where $\delta_{nm}$ is the Kronecker's delta.
Due to this normalized orthogonality, we obtain 
\tido{the} following relation\tido{s}:
\begin{align}
[\psi^{\dagger}_{n},\psi_{m}]_{+}&=\delta_{nm},\\
G_{IJ}=\langle c_{I}^{\dagger}c_{J}\rangle 
&=\frac{\langle \phi_{\rm SL}| c_{I}^{\dagger}c_{J} | \phi_{\rm SL}\rangle}{\langle \phi_{\rm SL}|\phi_{\rm SL}\rangle } \\
&=\sum_{n} \Phi_{In}^{*} \Phi_{Jn}.
\end{align}

Here, we rewrite $\ket{\phi_{\rm SL}}$ and obtain {an} explicit 
relation between $F_{IJ}$ and $\Phi_{In}$.
By using the commutation relation for $\psi^{\dagger}_{n}$,
we rewrite $\tido{\ket{\phi_{\rm SL}}}$ as 
\begin{align}
  |\phi_{\rm SL}\rangle \propto \prod_{n=1}^{N_{\tido{\rm e}}/2}\Big(\psi_{2n-1}^{\dagger}\psi_{2n}^{\dagger}\Big)|0\rangle.
\end{align}
From this, we obtain the following relations as
\begin{align}
|\phi_{\rm SL}\rangle &\propto \prod_{n=1}^{\Ne/2}\Big(\psi_{2n-1}^{\dagger}\psi_{2n}^{\dagger}\Big)|0\rangle
=\prod_{n=1}^{\Ne/2} K_{n}^{\dagger}|0\rangle \\
&\propto\Big(\sum_{n=1}^{{N_{\tido{\rm e}}}/{2}}K_{n}^{\dagger}\Big)^{{\Ne}/{2}} |0\rangle
=\Big(\sum_{I,J=0}^{2\Ns-1}\Big[\sum_{n=1}^{{\Ne}/{2}}{\Phi_{I 2n-1}\Phi_{J 2n}}\Big]
c_{I}^{\dagger}c_{J}^{\dagger}\Big){^{{\Ne}/{2}}}|0\rangle,
\end{align}
where $K_{n}^{\dagger}=\psi_{2n-1}^{\dagger}\psi_{2n}^{\dagger}$ and
we use the relation  $K_{n}^{\dagger}K_{m}^{\dagger}=K_{m}^{\dagger}K_{n}^{\dagger}$ {for $n\neq m$}.
This result shows that $F_{IJ}$ can be 
\tg{described} as
\begin{align}
  F_{IJ}=\sum_{n=1}^{{N_{\tido{\rm e}}}/{2}}\Big({\Phi_{I 2n-1}\Phi_{J 2n}-\Phi_{J 2n-1}\Phi_{I 2n}}\Big).
\end{align}
We note that this is one of expressions of $F_{IJ}$ for 
single Slater determinant, i.e, $F_{IJ}$ depend on
the pairing degrees of freedom {(how to make pairs from $\psi_{n}^{\dagger}$)} and
gauge degrees of freedom (we can {arbitrarily} change
the phase of $\Phi$ as $\Phi_{I}\rightarrow e^{{\rm i}\theta}\Phi_{I}$).
This large {number of} degrees of freedom is the origin of huge redundancy in $F_{IJ}$.

For 
\tido{$S^z=0$} case, i.e., $N_{\rm up}=N_{\rm down}$,
the anti-parallel Pfaffian wave functions and Slater determinant are given as 
\begin{align}
  &|\phi_{\rm A{P}-SL}\rangle=\Big(\prod_{n=1}^{N_{\tido{\rm e}}/2}\psi_{n\uparrow}^{\dagger}\Big)
  \Big(\prod_{m=1}^{N_{\tido{\rm e}}/2}\psi_{m\downarrow}^{\dagger}\Big)|0\rangle,\psi_{n\sigma}^{\dagger}=\sum_{i=0}^{N_{\tido{\rm s}}-1}\Phi_{in\sigma}c^{\dagger}_{i\sigma},\\
  &|\phi_{\rm A{P}-Pf}\rangle=\Big(\sum_{i,j=0}^{N_{\tido{\rm s}}-1}f_{ij}c_{i\uparrow}^{\dagger}c_{j\downarrow}^{\dagger}\Big)^{N_{\rm e}/2}|0\rangle.
\end{align}
By using the similar argument, we obtain the relation as follows:
\begin{align}
  f_{ij}=\sum_{n=1}^{{N_{\tido{\rm e}}}/{2}}\Phi_{in\uparrow}\Phi_{jn\downarrow}.
\end{align}

\section{Differentiation of Pfaffian}
\label{sec:diffPf}
We prove the relation given in Eq.~(\ref{Eq:DiffPf}).
To prove the relation, we use the following 
relations given as 
\begin{align}
&{\rm Pf}(BAB^{T})={\rm det}B\times{\rm Pf}A, \label{Eq:detPf}\\
&{\rm det}(I+B\delta x)=1+{\rm Tr}[B]\delta x+O(\delta x^2)\label{Eq:det}
\end{align}
where $A$ is a skew-symmetric matrix,
$B$ is a\tido{n} arbitrary square matrix,
and $\delta x$ denotes infinitesimal quantity.
A proof of Eq.~(\ref{Eq:detPf}) is shown in the literature~\cite{Tahara_JPSJ2008}
and it is easy to 
\tohgoe{prove} Eq.~(\ref{Eq:det}) by simple calculations.
It is possible to expand {a} Pfaffian with respect {to}  small $\delta x$ as follows:
\begin{align}
  &{\rm Pf}[A(x+\delta x)]={\rm Pf}[A(x)+\frac{\partial A(x)}{\partial x}\delta x+O(\delta x^2)] \\ \notag
&={\rm Pf}[(I+\frac{1}{2}A(x)^{-1}\frac{\partial A(x)}{\partial x}\delta x)^{T}A(x)(I+\frac{1}{2}A(x)^{-1}\frac{\partial A(x)}{\partial x}\delta x)+O(\delta x^2)] \\ \notag
&={\rm det}[I+\frac{1}{2}A(x)^{-1}\frac{\partial A(x)}{\partial x}\delta x+O(\delta x^2)]{\rm Pf}[A(x)] \\ \notag
&={\rm Pf}[A(x)](1+\frac{1}{2}{\rm Tr}[A(x)^{-1}\frac{\partial A(x)}{\partial x}]\delta x+O(\delta x^2)]) \tido{,}
\end{align}
From this, we prove Eq.~(\ref{Eq:DiffPf}) as follows:
\begin{align}
&\frac{\partial {\rm Pf}[A(x)]}{\partial x} 
=\lim_{\delta x\rightarrow 0} \frac{{\rm Pf}[A(x+\delta x)]-{\rm Pf}[A(x)]}{\delta x} \\
&=\lim_{\delta x\rightarrow 0} \frac{1}{2}\frac{{\rm Pf}[A(x)]{\rm Tr}[A(x)^{-1}\frac{\partial A(x)}{\partial x}]\delta x+O(\delta x^2)}{\delta x} \\
&=\frac{1}{2}{\rm Pf}[A(x)]{\rm Tr}\Big[A(x)^{-1}\frac{\partial A(x)}{\partial x}\tido{\Big].}
\end{align}

For simplicity, we consider the derivative with respect to
the real part of $F_{IJ}$ or $f_{ij}$.
In mVMC, because $(X)_{IJ}=F_{IJ}-F_{JI}$,
its derivative with respect to $F_{AB}^{R}$ is given by
\begin{align}
\Big[\frac{\partial{\rm Pf}(X)}{\partial F_{AB}^{R}}\Big]_{IJ}=\delta_{I,A}\delta_{J,B}-\delta_{J,A}\delta_{I,B}.
\end{align}
Thus, we obtain
\begin{align}
  {\rm Tr}\Big[X^{-1}\frac{\partial{\rm Pf}(X)}{\partial F_{AB}^{R}}\tido{\Big]}=-2(X^{-1})_{AB},
\end{align}
where we use the fact that $X^{-1}$ is also a skew-symmetric matrix.
In a similar way, 
for the spin-projected wave functions,
{the} derivative with respect to $f_{ab}^{R}$ is given by
\begin{align}
\Big[\frac{\partial{\rm Pf}(X)}{\partial f_{ab}^{R}}\Big]_{IJ}
=\delta_{i,a}\delta_{j,b}\kappa(\sigma,\sigma^{\prime})
-\delta_{i,b}\delta_{j,a}\kappa(\sigma^{\prime},\sigma),
\end{align}
and we obtain
\begin{align}
  {\rm Tr}\Big[X^{-1}\frac{\partial{\rm Pf}(X)}{\partial f_{ab}^{R}}\tido{\Big]}=-2\sum_{\sigma,\sigma^{\prime}}\kappa(\sigma,\sigma^{\prime})(X^{-1})_{a\sigma,b\sigma^{\prime}}.
\end{align}
Using the above relations, it is easy to obtain Eq.~(\ref{eq:diffij}).














\end{document}